\documentclass[a4paper,11pt]{article}
\usepackage{jheppub} 

\usepackage{caption}    
\usepackage{float}       
\usepackage{array}
\usepackage{booktabs}
\usepackage{multirow}
\usepackage[table]{xcolor}
\arxivnumber{} 

\title{\boldmath Superconformal index for $\N = 4$ Super Yang-Mills and  Elliptic Macdonald Polynomials  }
\preprint{
    USTC-ICTS/PCFT-26-23
}

\author{Gao-fu Ren}\author{and Min-xin Huang} 
\affiliation{Interdisciplinary Center for Theoretical Study, University of Science and Technology of China,  Hefei, Anhui 230026, China}
\affiliation{Peng Huanwu Center for Fundamental Theory, \\  Hefei, Anhui 230026, China}

\emailAdd{rengaofu@mail.ustc.edu.cn, minxin@ustc.edu.cn}

\abstract{We establish a connection between the superconformal index of $\N=4$ $U(N)$ SYM and the elliptic Ruijsenaars-Schneider integrable system. The index admits an expression in terms of elliptic Macdonald polynomials, which leads to a compact summation over generalized partitions involving the structure constants $B_\lambda(p,q,t)$ and normalization constants $\N_\lambda(p,q,t)$. By solving the elliptic Ruijsenaars-Schneider model perturbatively in the elliptic parameter $p$, a systematic expansion of the index in powers of $p$ is obtained. We check that in various limits, namely a deformed 1/2 BPS limit and  especially the large $N$ limit, our formalism reduces to previously known results. }

\begin{document}
\maketitle
\flushbottom

\section{Introduction}
The AdS/CFT correspondence establishes a powerful duality between certain gauge theories and theories of quantum gravity. One of its most concrete realizations is the duality between four-dimensional $\mathcal{N}=4$ supersymmetric Yang-Mills (SYM) theory with gauge group $U(N)$ and type IIB string theory on an $AdS_5 \times S^5$ background. A key quantitative tool for probing this duality is the superconformal index, defined as a refined Witten index in radial quantization \cite{Romelsberger:2005eg, kinney_2007}. Protected against quantum corrections, the index can be computed exactly at weak coupling and reliably extrapolated to strong coupling, where it becomes relevant to the dual gravity theory.

Superconformal indices have also been instrumental in counting microstates of supersymmetric black hole in $AdS_5$ space through the holographic dual 4d CFT \cite{cabo-bizet:2018ehj}. Specifically, for large $N$, the logarithm of the index should asymptotically reproduce the Bekenstein-Hawking entropy of supersymmetric AdS black holes. This connection has been rigorously verified in recent years through sophisticated asymptotic analyses of the index, such as the Cardy-like expansion \cite{choi:2018hmj} and the Bethe ansatz approach \cite{benini:2018ywd}. These novel developments have entailed the interesting problem of distinguishing the graviton
states and non-graviton states, or fortuitous states \cite{chang:2022mjp,choi:2023znd}. For some more recent discussions see  e.g.  \cite{Chen:2024oqv, Padellaro:2024rld, Chang:2024lxt,  Hughes:2020fpu, deMelloKoch:2025qeq,vanleuven:2025gwr}.

In our paper, we will consider the $\N = 4$ $U(N)$ SYM, maximally supersymmetric Yang-Mills (SYM) theory,
which is particular relevant for holographic black hole microstate counting, as well as tests of the giant graviton expansion proposed in \cite{arai:2020qaj,imamura:2021ytr,gaiotto:2021xce,murthy:2022ien}. The giant graviton expansion interprets the coefficients in the expansion of the finite $N$ superconformal index in a fugacity parameter as contributions of D-branes. However, the quantitative tests of the relations usually require analytic continuation of the fugacity parameters, so some closed formulas of the superconformal index are urgently needed. For some recent studies related to giant graviton expansion, see e.g. \cite{Jejjala:2022lrm, Deddo:2024liu, Hatsuda:2024uwt, Hayashi:2024aaf, Lee:2025veh}. 

In another perspective, the deep interplay between superconformal indices and integrable systems has emerged as a powerful theme in theoretical physics, particularly in the context of
$\N=2$ superconformal field theories of class $\mathcal{S}$ \cite{gadde:2011uv,gaiotto:2012xa,razamat:2013qfa,rastelli:2014jja}. These theories arise from compactifying the six-dimensional $(2,0)$ theory on a Riemann surface with punctures, and their structure is deeply constrained by S-duality. In this setting, the 4d/2d correspondence provides a powerful framework for constructing superconformal indices with maximal $A_N$-type flavor punctures. The construction proceeds by assembling fundamental building blocks, namely, the propagator and the three-punctured sphere, or trinion, via an integral transformation whose measure is dictated by the propagator. The following formula serves as the main motivation for our paper in these works, under the assumption of S-duality, the $A_1$ trinion admits an expansion of the form
\begin{align}\label{trinion}
I^{(0,3)}(a,b,c)=\sum_{\lambda}\mathcal{C}_{\lambda}(p,q,t)\psi_{\lambda}(a)\psi_{\lambda}(b)\psi_{\lambda}(c)\,,
\end{align}
where $\psi_{\lambda}(x)$ are eigenfunctions of the $A_1$ elliptic Ruijsenaars-Schneider model, defined up to a similarity transformation. This expansion encodes the precise way in which the index factorizes onto punctures and serves as a bridge to integrability. The generalization to higher-rank $A_N$ cases follows a similar pattern, with the trinion block playing an analogous role. Beyond the class $\mathcal{S}$ setting, further connections between elliptic integrable models and supersymmetric index of 6d SCFT have been explored in \cite{nazzal:2018brc,nazzal:2021tiu}, revealing a rich and unifying web of relations.

In this paper, we study a similar relation between the superconformal index of $\N=4$ $U(N)$ SYM and the elliptic Ruijsenaars-Schneider model and results are summarized as follows. We generalize the novel method of \cite{hatsuda2025deformedschurindicesmacdonald} to the case of full index, using elliptic Macdonald polynomials. We begin with the integral representation of the superconformal index for $\mathcal{N}=4$ $U(N)$ SYM in terms of an elliptic hypergeometric integral \cite{Spiridonov_2011}.  The integrand naturally factorizes into two distinct components: a weight function and a kernel function
\begin{align}\label{SCIweiker}
    I_{N} = \chi'_N \int_{\mathbb{T}^{N}} 
\underbrace{\left(\prod_{1 \leq i \neq j \leq N} \frac{\Gamma(t x_i/x_j; p, q)}{\Gamma( x_i/x_j; p, q)}\right)}_{\text{weight function}}\ \underbrace{\left(\prod_{1 \leq i , j \leq N}\frac{\Gamma(u x_i/x_j; p, q)}{\Gamma(tu x_i/x_j; p, q)} \right)}_{\text{kernel function }}
\prod_{j=1}^{N} 
\frac{dx_j}{2\pi i x_j}\,,
\end{align}
where the weight function plays the role of a measure. Up to a similarity transformation, it coincides precisely with the propagators appearing in the $\mathcal{N}=2$ class $\mathcal{S}$ construction. The kernel function, on the other hand, admits a diagonal expansion in terms of elliptic Macdonald polynomials, with the fugacity $u$ serving as the weight parameter:
\begin{align}\label{kernelf}
    \prod_{1 \leq i , j \leq N}\frac{\Gamma(u x_i/x_j; p, q)}{\Gamma(tu x_i/x_j; p, q)}=\sum_{\lambda\in \Lambda^N}u^{|\lambda|}B_{\lambda}(p,q,t)\p_\lambda(\x;p,q,t)\p_\lambda(\x^{-1};p,q,t)\,.
\end{align}
The sum runs over generalized partitions $\lambda\in \Lambda^N$, where $\Lambda^N=\{\lambda\in\mathbb{Z}^N\mid \lambda_1\geq \lambda_2\geq \cdots \geq\lambda_N\}$, and $\p_\lambda(\mathbf{x};p,q,t)$ denotes the eigenfunctions of the $A_{N-1}$ elliptic Ruijsenaars Schneider model on the physical domain $\mathbb{T}^N$, 
commonly referred to as elliptic Macdonald polynomials. Although strictly speaking they are not polynomial functions, we retain the conventional terminology. Under the same similarity transformation alluded to earlier, these elliptic Macdonald polynomials correspond exactly to the eigenfunctions $\psi_\lambda(\x)$ introduced in the $\mathcal{N}=2$ context. From this perspective, the $\mathcal{N}=4$ $U(N)$ SYM superconformal index can be interpreted as a $u$-weighted sphere with two punctures, effectively self-glued via the kernel expansion. A crucial property of the elliptic Macdonald polynomials is their orthogonality with respect to the weight function:
\begin{align}\label{orthog}
\frac{1}{N!}\int_{\mathbb{T}^{N}} \omega(\x;p,q,t),\p_\lambda(\x^{-1};p,q,t)\p_\mu(\x;p,q,t)\prod_{j=1}^{N}\frac{dx_j}{2\pi i x_j}
= \delta_{\lambda,\mu}\mathcal{N}_\lambda(p,q,t)\,,
\end{align}
where $\omega(\mathbf{x};p,q,t)$ is just the weight function appears in (\ref{SCIweiker}) and $\mathcal{N}_\lambda(p,q,t)$ denotes the squared norm.  This orthogonality allows us to simplify the index to a discrete sum over generalized integer partitions:
\begin{align}
    \label{SCIsumlam}
    I_N=\chi'_N\sum_{\lambda\in\Lambda^N}u^{|\lambda|}B_{\lambda}(p,q,t)\N_{\lambda}(p,q,t)\,.
\end{align}

It is useful to compare~\eqref{SCIsumlam} with the class $\mathcal S$ TQFT
construction. Self-gluing two punctures of the trinion (\ref{trinion}) gives,
in the normalization used here, the once-punctured torus amplitude
\begin{equation}
 I^{(1,1)}(a)
 =
 \sum_{\lambda}
 \mathcal{C}_\lambda(p,q,t)\N_\lambda(p,q,t)\psi_\lambda(a).
\end{equation}
Compared with~\eqref{SCIsumlam}, this expression contains one additional
wavefunction associated with the remaining puncture. Formally, one
may try to close this puncture by introducing a cap amplitude
\begin{equation}
 I^{(0,1)}(a)
 =
 \sum_{\lambda}
 D_\lambda(p,q,t)\psi_\lambda(a).
\end{equation}
Gluing the two amplitudes would then give, up to overall factors and
normalization conventions,
\begin{equation}
 I^{(1,0)}_{\mathrm{formal}}
 \sim
 \sum_{\lambda}
 \mathcal{C}_\lambda D_\lambda N_\lambda^2.
\end{equation}
Comparison with~\eqref{SCIsumlam} suggests that the presently unknown coefficients
$D_\lambda$ could provide a relation between the class $\mathcal S$
coefficients $C_\lambda$ and our kernel coefficients $B_\lambda$.

This interpretation should, however, be regarded as heuristic. The
standard class $\mathcal S$ construction is formulated for stable
surfaces satisfying $2g-2+n>0$, whereas the unpunctured torus
$(g,n)=(1,0)$ and the cap $(g,n)=(0,1)$ do not satisfy this condition.
Moreover, from an $\mathcal N=2$ viewpoint, the additional fugacity
$u$ may be interpreted as weighting a combination of the adjoint
hypermultiplet flavor charge and the R-symmetry charges, so the
hypothetical coefficients $D_\lambda$ may naturally depend on $u$.
Whether such a cap amplitude can be defined intrinsically, and whether
it yields a precise relation between $C_\lambda$ and $B_\lambda$, are
interesting questions that we leave for future consideration.

In the limit $p\to 0$, the superconformal index reduces to the deformed Schur index, for which an analogous expansion in terms of ordinary Macdonald polynomials has been established \cite{hatsuda2025deformedschurindicesmacdonald}. In this limit, the fundamental structure constants admit explicit closed-form expressions. For the full elliptic case (\ref{SCIsumlam}), however, explicit formulas for $B_{\lambda}(p,q,t)$ and $\N_{\lambda}(p,q,t)$ remain unknown. Nevertheless, these quantities can be computed order by order as formal power series in $p$, yielding a systematic series expansion of the superconformal index in $p$. In addition, using the symmetric function approach, we obtain the well known exact expression for the index in the large $N$ limit.

This paper is organized as follows. In Section \ref{sec2}, we introduce the elliptic Ruijsenaars-Schneider model and present a systematic method for solving it perturbatively in the parameter $p$. This approach yields an expansion of the elliptic Macdonald polynomials in terms of generalized monomial symmetric functions, and we state two key theorems that will be essential for the subsequent analysis. In Section \ref{sec3}, we introduce the superconformal index of $\mathcal{N}=4$ $U(N)$ SYM. Starting from its elliptic hypergeometric integral representation, we show that the index can be rewritten in the form (\ref{SCIsumlam}). In Section \ref{sec4}, we interpret the full superconformal index as an elliptic lift of the deformed Schur index \cite{hatsuda2025deformedschurindicesmacdonald}, with $p$ playing the role of the elliptic deformation parameter and treating this lift perturbatively. Then we consider the limit $q=t$, and finally according to symmetric function theory we obtain the well known exact expression in the large $N$ limit. In Section \ref{sec5}, we compute the elliptic normalization constants $\N_\lambda(p,q,t)$ and the structure constants $B_\lambda(p,q,t)$ perturbatively by solving the elliptic Ruijsenaars Schneider model, and we evaluate the superconformal index for the $N=2$ case. In Section \ref{sec6}, we conclude by summarizing our results and outlining several promising directions for future research. In Appendix \ref{appendixA}, we collect explicit results for the expansion of elliptic Macdonald polynomials in terms of generalized monomial symmetric functions.

\section{Elliptic Ruijsenaars Schneider model and Elliptic Macdonald Polynomials}\label{sec2}
In this section, we begin by introducing the elliptic Ruijsenaars Schneider model, a well known classical integrable system of elliptic type. For each $r = 0, 1,\dots, N$, we define the $r$th order $q$-difference operator $\mathcal{D}^{(r)}_x(p) := \mathcal{D}^{(r)}_x(p|q,t)$
\begin{align}
    \mathcal{D}^{(r)}_x(p)=\Bigg(t^{\binom{r}{2}}\Bigg)\sum_{\substack{I\subseteq \{1,\dots,N\} \\ |I|=r}}\prod_{\substack{i\in I\\j \notin I}}\frac{\theta(tx_i/x_j;p)}{\theta(x_i/x_j;p)}\prod_{i\in I}T_{i}\,,
\end{align}
where $T_{i}$ denotes the $q$-shift operator acting on the coordinate $x_i$ as$ $ $T_{i}f(x_1,\dots,x_i,\dots,x_N)$ $=f(x_1,\dots,qx_i,\dots,x_N)$. In this work, we are primarily interested in the expansion in the parameter $p$. Accordingly, we express each of these $q$-difference operators as a formal power series in $p$ of the form
\begin{align}
    \mathcal{D}^{(r)}_x(p)=\sum_{k=0}^\infty p^k\mathcal{D}^{(r)}_{x,k}\,.
\end{align}
These operators commute with one another and satisfy a set of joint eigenvalue equations:
\begin{align}
    \mathcal{D}^{(r)}_x(p)\psi(\mathbf{x};p)=\varepsilon^{(r)}(p)\psi(\mathbf{x};p)\quad\quad\quad\quad (r=\pm1,\dots,\pm N)\,,
\end{align}
where the operators with negative indices are defined through the relation
\begin{align}
    \mathcal{D}^{(-r)}_x(p)=\mathcal{D}^{(N-r)}_x(p)\mathcal{D}^{(N)}_x(p)^{-1}\,, 
\end{align}
On the physical domain $\mathbf{x}\in\mathbb{T}^N$, the common eigenfunctions of this commuting family are given by the so-called elliptic Macdonald polynomials. Because we are primarily concerned with their expansion in the parameter $p$, we should express both the eigenfunctions and the corresponding eigenvalues as formal power series in $p$. The joint eigenfunctions on the physical torus can be viewed as elliptic deformations of the ordinary Macdonald polynomials $P_\lambda(\mathbf{x};q,t)$. Although strictly speaking they are not polynomial functions, we adhere to the conventional terminology and continue to refer to them as elliptic Macdonald polynomials. These formal eigenfunctions admit a series representation of the form \cite{mr4405564}
\begin{align}
    \mathcal{P}_\lambda(\mathbf{x};p,q,t)=\sum_{k=0}^\infty p^k\mathcal{P}_{\lambda,k}(\mathbf{x};q,t)\,,
\end{align}
where the leading term $\p_{\lambda,0}(\mathbf{x};q,t)$ coincides precisely with the ordinary Macdonald polynomial, i.e. the elliptic Macdonald polynomials reduce to ordinary Macdonald polynomials in the $p\to 0$ limit.

To normalize these eigenfunctions, we introduce the generalized monomial symmetric functions. For a generalized partition $\lambda\in \Lambda^N$, where $\Lambda^N=\{\lambda\in\mathbb{Z}^N\mid \lambda_1\geq \lambda_2\geq \cdots \geq\lambda_N\}$ denotes the set of generalized integral partitions. Unlike the usual partitions, here the $\lambda_i$'s can be negative. We define
\begin{align}
    m_{\lambda}(\mathbf{x})=\sum_{\mu\in\mathfrak{S}_n \cdot \lambda}x^\mu\,,
\end{align}
where the sum runs over all distinct monomials in the orbit of $\lambda$ under the symmetric group $\mathfrak{S}_n$, and $x^\mu=x_1^{\mu_1}\cdots x_N^{\mu_N}$ for $\mu=(\mu_1,\ldots,\mu_N)\in \mathbb{Z}^N$. A formal solution $\p_\lambda(\mathbf{x};p,q,t)$ is said to be normalized if, for all $k>0$, the coefficient of $m_{\lambda}(\mathbf{x})$ in $\p_{\lambda,k}(\mathbf{x};q,t)$ vanishes. More concretely, this normalization condition takes the form proposed in \cite{mr4405564}:
\begin{align}\label{normization}
    \mathcal{P}_{\lambda,k}(\mathbf{x};q,t)=\sum_{\mu\leq \lambda+k\phi}C^k_{\lambda\mu}(q,t)m_\mu(\mathbf{x}),\quad\quad\quad C^k_{\lambda\lambda}=\delta_{k,0}\quad\quad (k=0,1,2,\dots)\,,
\end{align}
where $\phi=(1,0,\ldots,0,-1)$ and $\mu\leq \nu$ denotes the dominance order on $\mathbb{Z}^N$, defined by the requirements
\begin{align}
    |\mu|=|\nu|\quad \text{and} \quad \mu_1+\cdots+\mu_i\leq \nu_1+\cdots+\nu_i \quad\quad  (i=1,\dots,N)\,.
\end{align}
The condition $\mu\leq \lambda+k\phi$ therefore implies that each component $\mu_j$ lies within the range $\lambda_N-k\leq \mu_j\leq\lambda_1+k$. Consequently, for fixed $\lambda$, the sum $\sum_{\mu\leq \lambda+k\phi}$ includes an increasing number of terms as $k$ grows.
All coefficients $C_{\lambda\mu}^k(q,t)$ can in principle be determined recursively by solving the eigenvalue problem order by order in perturbation theory.

Although explicit closed form formulas for the elliptic Macdonald polynomials have been conjectured in the literature \cite{mr4040584}, they will not be needed in the present work.

The elliptic deformation of the weight function appearing in the orthogonal relation for ordinary Macdonald polynomials is given by
\begin{align}\label{weif}
    \omega(\mathbf{x};p,q,t)=\prod_{1 \leq i < j \leq N} \frac{\Gamma(t (x_i/x_j)^{\pm 1}; p, q)}{\Gamma( (x_i/x_j)^{\pm 1}; p, q)}\,,
\end{align}
With respect to this measure, the elliptic Macdonald polynomials satisfy an orthogonality relation \cite{mr4405564}, which we state as the following theorem.
\begin{theorem}\label{thm1}
Suppose that $|t|< 1$ and $t^
k \notin q^{\mathbb{Z}>0}$,$(k = 1,\dots,n - 1)$. Then, the normalized joint eigenfunctions $\p_\lambda(\mathbf{x}; p,q,t)$ of the elliptic Ruijsenaars operators, attached to the dominant vectors $\lambda$ in
$\mathbb{Z}^N$, are orthogonal with respect to the weight function $\omega(\mathbf{z};p,q,t)$ in the sense that
    \begin{align}
    \label{ortho}
    \langle \p_\lambda(\mathbf{x})|\p_\mu(\mathbf{x}) \rangle:=\frac{1}{N!}\int_{\mathbb{T}^{N}} \omega(\mathbf{x};p,q,t)\p_\lambda(\mathbf{x}^{-1})\p_\mu(\mathbf{x})\prod_{j=1}^{N} \frac{dx_j}{2\pi i x_j}=\delta_{\lambda,\mu}\mathcal{N}_\lambda(p,q,t)\,.
\end{align}
\end{theorem}
\noindent 
Here $\N_\lambda(p,q,t)$ denotes the elliptic generalization of the normalization constant $\N_\lambda(q,t)$ for ordinary Macdonald polynomials. The technical conditions $t^k\notin q^{\mathbb{Z}_{>0}}$ $(k=1,\dots,n-1)$ are expected to be removable.

The kernel function associated with this integrable system is defined as
\begin{align}
    \label{kerf}
    K_u(\mathbf{x},\mathbf{y})=\prod_{1 \leq i,j \leq N}\frac{\Gamma(ux_iy_j; p, q)}{\Gamma(tux_iy_j; p, q)}\,,
\end{align}
which satisfies the symmetry property
\begin{align}
    \mathcal{D}^{(r)}_x(p)K_u(\mathbf{x},\mathbf{y})= \mathcal{D}^{(r)}_y(p)K_u(\mathbf{x},\mathbf{y})\,.
\end{align}
To derive the key result (\ref{SCIsumlam}), we rely on the following theorem \cite{rains:2025fqe}:
\begin{theorem}\label{thm2}
\begin{align}\label{ellpcauchy}
    K_u(\mathbf{x},\mathbf{y})=\sum_{\lambda\in\Lambda^N}u^{|\lambda|}B_{\lambda}(p,q,t)\p_{\lambda}(\mathbf{x};p,q,t)\p_{\lambda}(\mathbf{y};p,q,t)\,.
\end{align}
\end{theorem}
\noindent In the non-elliptic limit $p\to 0$, the elliptic Macdonald polynomials reduce to ordinary Macdonald polynomials. Correspondingly, the normalization constants $\N_\lambda(p,q,t)$ and the structure constants $B_\lambda(p,q,t)$ reduce to their non-elliptic counterparts, which admit explicit closed formulas:
\begin{align}\label{blambdaandNormalconst}
\begin{split}
    \N_{\lambda,N}(q,t)&=\prod_{1 \leq i<j \leq N} \frac{(t^{j-i}q^{\lambda_i-\lambda_j+1};q)_\infty (t^{j-i}q^{\lambda_i-\lambda_j};q)_\infty}{(t^{j-i+1}q^{\lambda_i-\lambda_j};q)_\infty (t^{j-i-1}q^{\lambda_i-\lambda_j+1};q)_\infty}\,,\\
    b_\lambda(q,t)&=\prod_{1 \leq i \leq j \leq \ell(\lambda)} \frac{(t^{j-i+1}q^{\lambda_i-\lambda_j};q)_{\lambda_j-\lambda_{j+1}}}{(t^{j-i}q^{\lambda_i-\lambda_j+1};q)_{\lambda_j-\lambda_{j+1}}}\,.
\end{split}
\end{align}

\section{Superconformal index of $\mathcal{N}=4$ $U(N)$ SYM}\label{sec3}
We begin with the matrix integral representation of the $\mathcal{N}=4$ superconformal index for the $U(N)$ gauge group. As established in \cite{kinney_2007}, after adjusting the fugacities to match our notation, the index takes the form
\begin{equation}
\begin{aligned}
I_N(t, u, v;p, q)=\int_{U(N)} dU  \exp \left( \sum_{n=1}^\infty \frac{f(t^n, u^n, v^n; p^n, q^n)}{n} \mathrm{Tr} (U^n) \mathrm{Tr} [ (U^\dagger)^n ] \right),
\label{eq:SCI}
\end{aligned}
\end{equation}
where $f(t, u, v;p, q)$ denotes the single letter index of the theory, expressed as
\begin{align}\label{eq:single-letter-index}
f(t, u, v; p, q)=1-\frac{(1-t)(1-u)(1-v)}{(1-p)(1-q)}\,.
\end{align}
It should be noted that the five parameters $(t, u, v; p, q)$ are not all independent, they satisfy the constraint $pq=tuv$, which leaves four independent degrees of freedom.

Since the integrand in \eqref{eq:SCI} is a class function of the unitary matrix $U$, we may apply Weyl's integration formula to reduce the integral to the maximal torus $\mathbb{T}^N$. This yields
\begin{equation}
\begin{aligned}
I_N(t, u, v; p, q)&=\frac{1}{N!} \oint_{\mathbb{T}^N} \prod_{i=1}^N \frac{dx_i}{2\pi i x_i} \prod_{1 \leq i \ne j \leq N} \left( 1-\frac{x_i}{x_j} \right)\\
&\quad \times \exp \left( \sum_{n=1}^\infty \frac{f(t^n, u^n, v^n; p^n, q^n)}{n} \powsym_n(\x) \powsym_n(\x^{-1}) \right),
\end{aligned}
\label{eq:SCI-2}
\end{equation}
where $\x=(x_1,\dots, x_N)$ are the eigenvalues of $U$, and $\powsym_n(\x)$ denotes the power sum symmetric polynomial. The integration contour for each $x_i$ is taken counterclockwise around the unit circle.

To obtain the elliptic hypergeometric integral representation of the superconformal index for $\N=4$ $U(N)$ SYM, we require several key identities. First, recall the following useful relations:
\begin{align}
    \prod_{1 \leq i \ne j \leq N} \left( 1-\frac{x_i}{x_j} \right)=\mathrm{PE}\left(\sum_{\alpha\in\Delta}\x^\alpha\right)^{-1}=\mathrm{PE}\left(-\sum_{\alpha\in\Delta}\x^\alpha\right),\  \powsym_n(\x) \powsym_n(\x^{-1})=\sum_{\alpha}\x^{n\alpha}+N\,,
\end{align}
as well as the exponential representation of the elliptic gamma function
\begin{align}\label{ellgammadef}
    \Gamma(x;p,q)=\e\bigg[\sum_{n=1}^\infty\frac{x^n-(pq)^nx^{-n}}{n(1-p^n)(1-q^n)}\bigg]\,,
\end{align}
Employing these identities, the integrand in \eqref{eq:SCI-2} can be rearranged into the form
\begin{align}
   \mathrm{PE}\left(f(t,u,v;p,q)(\chi_{adj}(\x)+N)-\sum_{\alpha\in\Delta}\x^\alpha\right)\,,
\end{align}
where $\chi_{adj}(\x)=\sum_{1\leq i\neq j\leq N}x_i/x_j=\sum_{\alpha\in\Delta}\x^\alpha$. To avoid potential divergences, we reorganize the terms as follows:
\begin{align}
    \exp&\bigg[\sum_{n=1}^\infty\bigg(\frac{t^n-(pq/t)^n-(tu)^n+(pq/tu)^n+u^n-(pq/u)^n}{n(1-p^n)(1-q^n)}(\chi_{adj}(\x^n)+N)\notag\\
    &\quad\quad\quad\quad-\frac{1-(pq)^n}{n(1-p^n)(1-q^n)}\chi_{adj}(\x^n)-\frac{p^n}{n(1-p^n)}N-\frac{q^n}{n(1-q^n)}N\bigg)\bigg]\notag\\
    &=\frac{(p;p)^N_\infty(q;q)^N_\infty\Gamma^{N}(t; p, q)\Gamma^{N}(u; p, q)}{\Gamma^{N}(tu; p, q)}\prod_{1 \leq i < j \leq N} \frac{\Gamma(t (x_i/x_j)^{\pm 1}; p, q)}{\Gamma( (x_i/x_j)^{\pm 1}; p, q)}\frac{\Gamma(u (x_i/x_j)^{\pm 1}; p, q)}{\Gamma(tu (x_i/x_j)^{\pm 1}; p, q)}\,.
\end{align}
Using the above manipulations, the superconformal index for $\N=4$ $U(N)$ SYM given in \eqref{eq:SCI-2} can be recast into the compact form
\begin{align}
\label{U(N)SCI3}
&I_{N} = \chi_N \int_{\mathbb{T}^{N}} 
\prod_{1 \leq i < j \leq N} \frac{\Gamma(t (x_i/x_j)^{\pm 1}; p, q)}{\Gamma( (x_i/x_j)^{\pm 1}; p, q)}\frac{\Gamma(u (x_i/x_j)^{\pm 1}; p, q)}{\Gamma(tu (x_i/x_j)^{\pm 1}; p, q)} 
\prod_{j=1}^{N} 
\frac{dx_j}{2\pi i x_j}\,,
\end{align}
with the prefactor
\begin{align}
    \chi_N=\frac{(p;p)^N_\infty(q;q)^N_\infty\Gamma^{N}(t; p, q)\Gamma^{N}(u; p, q)}{N!\Gamma^{N}(tu; p, q)}\,.
\end{align}
Alternatively, in terms of the elliptic weight function \eqref{weif} and the kernel function defined in \eqref{kerf} (with the identification $\mathbf{y}\to\mathbf{x}^{-1}$), the index simplifies to
\begin{align}
    \label{SCIusingkerf}
    I_N=\chi'_N\int_{\mathbb{T}^{N}} 
\omega(\x)K_u(\x,\x^{-1}) 
\prod_{j=1}^{N} 
\frac{dx_j}{2\pi i x_j}\,,
\end{align}
where
\begin{align}
    \chi'_N=\frac{(p;p)^N_\infty(q;q)^N_\infty\Gamma^{N}(t; p, q)}{N!}\,.
\end{align}
Now, applying Theorem \ref{thm1} and Theorem \ref{thm2}, we can expand the kernel function in terms of elliptic Macdonald polynomials and exploit their orthogonality. This leads to a discrete sum representation of the index:
\begin{align}\label{SCISUMLAM}
\begin{split}
    I_{N} &=\chi'_N\sum_{\lambda\in\Lambda^N}u^{|\lambda|}B_{\lambda}(p,q,t) \int_{\mathbb{T}^{N}} 
\omega(\z)\p_\lambda(\x;p,q,t) \p_\lambda(\x^{-1};p,q,t) 
\prod_{j=1}^{N} 
\frac{dx_j}{2\pi i x_j}\\
&=\chi'_N \sum_{\lambda\in \Lambda^N}u^{|\lambda|}B_{\lambda}(p,q,t)\mathcal{N}_\lambda(p,q,t)\,,
\end{split}
\end{align}
which coincides with the expression \eqref{SCIsumlam} anticipated in the introduction.

For the purpose of extracting the $u$-expansion, it is useful to decompose the kernel function as
\begin{align}
    K_u(\mathbf{x},\mathbf{y})=\sum_{m=-\infty}^\infty K^{(m)}(\mathbf{x},\mathbf{y})u^m\,,
\end{align}
where each $K^{(m)}(\x,\y)$ admits an expansion over partitions of fixed weight: 
\begin{align}\label{Kmexpans}
    K^{(m)}(\mathbf{x},\mathbf{y})=\sum_{\lambda\in\Lambda^N,|\lambda|=m}B_{\lambda}(p,q,t)\p_{\lambda}(\mathbf{x};p,q,t)\p_{\lambda}(\mathbf{y};p,q,t)\,.
\end{align}
Here $|\lambda|=\lambda_1+\lambda_2+\cdots+\lambda_N$ denotes the total weight of the generalized integer partition. Consequently, the superconformal index can be expressed as a formal power series in the fugacity $u$:
\begin{align}
    \label{SCIinuexpansion}
    I_N(p,q,t,u)=\sum_{k\in \mathbb{Z}}u^kI^k_N(p,q,t)\,,
\end{align}
with coefficients $I^k_N(p,q,t)$ independent of $u$ and given by a sum over weight $k$ generalized integer partitions:
\begin{align}
    \label{I^k_N(p,q,t)}
    I^k_N(p,q,t)=\chi'_N\sum_{\lambda\in\Lambda^N,|\lambda|=k}B_{\lambda}(p,q,t)\mathcal{N}_\lambda(p,q,t)\,.
\end{align}
This provides a systematic way to compute the index order by order in $u$, once the elliptic data $B_\lambda(p,q,t)$ and $\N_\lambda(p,q,t)$ are determined.

\section{Perturbative expansion of the  SCI in  elliptic parameter $p$}\label{sec4}
 \subsection{Superconformal index in perturbation}
 In this subsection, we consider the total superconformal index as an elliptic lift of deformed Schur index. By interpreting the elliptic lift as a perturbation, we observe that the integrand in \eqref{U(N)SCI3} can be reorganized into the following form up to an overall factor:
\begin{align}\label{pexpaninmacform}
    \prod_{1\leq i \neq j \leq N}\frac{(x_i/x_j;q)_\infty}{(tx_i/x_j;q)_\infty}\prod_{i,j=1}^N\frac{(tux_i/x_j;q)_\infty}{(ux_i/x_j;q)_\infty}\cdot \Phi(\mathbf{x},p,q,t,u)\,,
\end{align}
where the first factor $\prod_{1\leq i \neq j \leq N}\frac{(x_i/x_j;q)_\infty}{(tx_i/x_j;q)_\infty}$ is precisely the weight function associated with Macdonald polynomials. The second factor $\prod_{i,j=1}^N\frac{(tux_i/x_j;q)_\infty}{(ux_i/x_j;q)_\infty}$ corresponds to the Macdonald type Cauchy kernel. The remaining factor $\Phi(\mathbf{x},p,q,t,u)$ is a formal power series in $p$ whose coefficients are Laurent polynomials symmetric in $\mathbf{x}$:
\begin{align}
    \Phi(\mathbf{x},p,q,t,u)=\sum_{k=0}^{\infty}p^k\Phi_k(\mathbf{x},q,t,u)\,.
\end{align}
To determine the explicit form of $\Phi(\mathbf{x},p,q,t,u)$, we examine the $p\to 0$ limit using the identity
\begin{align}
    \label{qpochexp}
    \Gamma(x;0,q)=\frac{1}{(x,q)_\infty}=\e\bigg[\sum_{n=1}^\infty\frac{x^n}{n(1-q^n)}\bigg]\,.
\end{align}
Applying \eqref{ellgammadef}, we can express the elliptic kernel function as
\begin{align}\label{ellkerfinexp}
   \prod_{1 \leq i,j \leq N}\frac{\Gamma(ux_i/x_j; p, q)}{\Gamma(tux_i/x_j; p, q)}=\e\bigg[\sum_{n=1}^\infty\frac{\left[u^n-(tu)^n-(pq/u)^n+(pq/tu)^n\right]\powsym_n(\x)\powsym_n(\x^{-1})}{n(1-p^n)(1-q^n)}\bigg]\,,
\end{align}
and similarly, the weight function combined with an overall factor takes the exponential form
\begin{align}\label{ellweifinexp}
    \begin{split}
\bigg((p;p)_\infty(q;q)_\infty\Gamma(t;p,q)\bigg)^N&\prod_{1 \leq i\neq j \leq N}\frac{\Gamma(tx_i/x_j;p,q)}{\Gamma(x_i/x_j;p,q)}\\
    &=\e\bigg[\sum_{n=1}^\infty\frac{\left[t^n-1-(pq/t)^n+(pq)^n\right]\powsym_n(\x)\powsym_n(\x^{-1})}{n(1-p^n)(1-q^n)}\bigg]\,.
    \end{split}
\end{align}
Consequently, the superconformal index can be recast as
\begin{align}\label{SCIellexprestomac}
    I_N=\chi''_N\int_{\mathbb{T}^{N}}\prod_{i=1}^N\frac{\mathrm{d}x_i}{2\pi \mathrm{i}x_i}\prod_{1\leq i \neq j \leq N}\frac{(x_i/x_j;q)_\infty}{(tx_i/x_j;q)_\infty}\prod_{i,j=1}^N\frac{(tux_i/x_j;q)_\infty}{(ux_i/x_j;q)_\infty}\cdot \Phi(\mathbf{x};p,q,t,u)\,,
\end{align}
with the prefactor $\chi''_N=\frac{1}{N!}\frac{(q;q)_\infty^N}{(t;q)_\infty^N}$. And the explicit expression for $\Phi(\mathbf{x};p,q,t,u)$ is given by
\begin{align}
\begin{split}
    \label{Phi}
    \Phi(\x;p,q,t,u)=\e\bigg[\sum_{n=1}^\infty \frac{(u^n-1)(1-t^n)p^n+(pq)^n(1-t^{-n})(1-u^{-n})}{n(1-p^n)(1-q^n)}\powsym_n(\x)\powsym_n(\x^{-1})\bigg]\,,
    \end{split}
\end{align}
which manifestly satisfies $\Phi(\x;0,q,t,u)=1$. This expression can be further expanded as
\begin{align}
    \label{levelp^n inPhi}
    \sum_{\Vec{m}}\frac{1}{1^{m_1}m_1!2^{m_2}m_2!\cdots }g_1^{m_1}g_2^{m_2}\cdots \powsym_1^{m_1}\powsym_2^{m_2}\cdots(\x)\powsym_1^{m_1}\powsym_2^{m_2}\cdots(\x^{-1})=\sum_{\mu}\frac{g_\mu}{z_\mu}\powsym_\mu(\x)\powsym_\mu(\x^{-1})\,,
\end{align}
where the summation runs over all tuples $(m_1,m_2,\dots)\in \mathbb{N}^\infty$. In the frequency representation, a partition $\lambda$ is denoted as $\lambda=(1^{m_1},2^{m_2},\cdots)$, indicating that the part $i$ appears $m_i$ times. The coefficients $g_i$ are defined as
\begin{align}
    \label{f_i}
    g_i=\frac{(u^i-1)(1-t^i)+q^i(1-t^{-i})(1-u^{-i})}{(1-q^i)(1-p^i)}p^i\,,
\end{align}
and we adopt the conventions
\begin{align}\label{defofzmupmu}
    g_\mu=\prod_{i=1}^{\ell(\mu)}g_{\mu_i}, \quad\quad\quad \powsym_\mu(\x)=\prod_{i=1}^{\ell(\mu)}\powsym_{\mu_i}(\x),\quad\quad\quad z_\mu=\prod_{i\geq1}i^{m_i}m_i!\,.
\end{align}
Introducing the auxiliary quantities
\begin{align}
    \label{h_i}
    \mathfrak{h}_i=\frac{(u^i-1)(1-t^i)+q^i(1-t^{-i})(1-u^{-i})}{(1-q^i)}\,,
\end{align}
the expansion \eqref{levelp^n inPhi} can be rewritten as
\begin{align}
    \label{Phiinh_i}
    \sum_{\mu}\frac{\mathfrak{h}_\mu}{z_\mu}\powsym_\mu(\x)\powsym_\mu(\x^{-1})\prod_{i=1}^{\ell(\mu)}\frac{p^{\mu_i}}{1-p^{\mu_i}}\,.
\end{align}
Expanding the factor $\prod_{i=1}^{\ell(\mu)}\frac{p^{\mu_i}}{1-p^{\mu_i}}$ as a power series in $p$ yields
\begin{align}
    \prod_{i=1}^{\ell(\mu)}\frac{p^{\mu_i}}{1-p^{\mu_i}}=p^{|\mu|}\sum_{r_1,\dots, r_{\ell(\mu)}=0}^{\infty}p^{\sum_{i=1}^{\ell(\mu)}r_i \mu_i}\,.
\end{align}
Let $\mathfrak{c}_n(\mu)$ denote the number of $\ell(\mu)$-tuples $(r_1,\cdots,r_{\ell(\mu)})$ satisfying $\sum_{i=1}^{\ell(\mu)}r_i \mu_i=n$. One can verify that $\mathfrak{c}_0(\mu)=1$. Writing a partition in frequency notation as $\lambda=(1^{k_1}2^{k_2}\dots)$, we obtain the following combinatorial expressions:
\begin{align}
\begin{split}
&\mathfrak{c}_1(\mu)=k_1,\quad \mathfrak{c}_2(\mu)=k_2+\frac{k_1\cdot(k_1-1)}{2}+k_1\,,\\
&\mathfrak{c}_3(\mu)=k_3+k_2\cdot k_1+\frac{k_1(k_1-1)(k_1-2)}{3!}+k_1\cdot (k_1-1)+k_1\,,\\
&\mathfrak{c}_4(\mu)=k_4+k_3\cdot k_1+k_2+\frac{k_2(k_2-1)}{2}+k_2\cdot k_1+\frac{k_2\cdot k_1(k_1-1)}{2}+\frac{k_1(k_1-1)(k_1-2)(k_1-3)}{4!}\\
&\quad\quad\ +k_1(k_1-1)+\frac{k_1(k_1-1)}{2}+k_1,\\
&\cdots\,.
\end{split}
\end{align}
Using these results, \eqref{Phiinh_i} can be expressed as a double series in $p$:
\begin{align}\label{pexph_i}
    \sum_{\mu}\sum_{n=0}^\infty\mathfrak{c}_n(\mu)p^{|\mu|+n}\frac{\mathfrak{h}_\mu}{z_\mu}\powsym_\mu(\x)\powsym_\mu(\x^{-1})\,.
\end{align}
For illustration, we present the explicit expansion of $\Phi(\x;p,q,t,u)$ up to order $p^2$. At order $p^1$, we find
\begin{align}
    \label{Phiinp^1}
    u^{-1}\frac{q(1-t)}{t(1-q)}\powsym_1(\x)\powsym_1(\x^{-1})+\frac{(q+t)(t-1)}{(1-q)t}\powsym_1(\x)\powsym_1(\x^{-1})+u\frac{1-t}{1-q}\powsym_1(\x)\powsym_1(\x^{-1})\,,
\end{align}
while at order $p^2$, the expression is given by
\begin{align}
    \begin{split}
        \label{Phiinp^2}
        &\quad\ u^{-2}\bigg[\frac{q^2(1-t^{-1})^2}{2(1-q)^2}\powsym^2_1(\x)\powsym^2_1(\x^{-1})+\frac{q^2(t^{-2}-1)}{2(1-q^2)}\powsym_2(\x)\powsym_2(\x^{-1})\bigg]\\
        &+u^{-1}\bigg[\frac{q(1-t)(1-t^{-1})-q^2(1-t^{-1})^2}{(1-q)^2}\powsym^2_1(\x)\powsym^2_1(\x^{-1})+\frac{q(t^{-1}-1)}{1-q}\powsym_1(\x)\powsym_1(\x^{-1})\bigg]\\
        &+u^0\bigg[\scalebox{1.1}{$\frac{(1-t)^2(1+(
        \frac{q}{t}
        )^2+\frac{4q}{t})}{2(1-q)^2}$}\powsym^2_1(\x)\powsym^2_1(\x^{-1})+\scalebox{1.1}{$\frac{(t-1)(1+\frac{q}{t})}{1-q}$}\powsym_1(\x)\powsym_1(\x^{-1})+\scalebox{1.1}{$\frac{(t^2-1)(1+\frac{q^2}{t^2})}{2(1-q^2)}$}\powsym_2(\x)\powsym_2(\x^{-1})\bigg]\\
        &+u^1\bigg[-\frac{((q/t)+1)(1-t)^2}{(1-q)^2}\powsym^2_1(\x)\powsym^2_1(\x^{-1})+\frac{1-t}{1-q}\powsym_1(\x)\powsym_1(\x^{-1})\bigg]\\
        &+u^2\bigg[\frac{(1-t)^2}{2(1-q)^2}\powsym^2_1(\x)\powsym^2_1(\x^{-1})+\frac{1-t^2}{2(1-q^2)}\powsym_2(\x)\powsym_2(\x^{-1})\bigg]\,.
    \end{split}
\end{align}
These explicit expansions provide the necessary ingredients for computing the superconformal index perturbatively in $p$
\begin{align}
    \label{I_N,k}
    I_N(p,q;t,u)=\sum_{k=0}^\infty p^kI_{N,k}(q;t,u)\,.
\end{align}
With the Macdonald Cauchy identity:
\begin{align}
    \label{MacCauchy}
    \prod_{i,j=1}^N \frac{(tu x_i/x_j;q)_\infty}{(ux_i/x_j;q)_\infty}=\sum_{\ell(\lambda) \leq N}u^{|\lambda|} b_\lambda(q,t) P_\lambda(\x;q,t)P_\lambda(\x^{-1};q,t)\,,
\end{align}
the superconformal index of $\mathcal{N}=4$ $U(N)$ SYM can be expressed as a power series in $p$. Substituting the expansion of $\Phi(\mathbf{x};p,q,t,u)$ derived in the previous, we obtain the $k$-th order coefficient $I_{N,k}$ in the $p$-expansion as
\begin{align}
    \label{I_N,Kintegralpexpansion}
    \begin{split}I_{N,k}&=\chi''_N\sum_{\ell(\lambda) \leq N} \sum_{|\mu|\leq k}u^{|\lambda|}\mathfrak{c}_{k-|\mu|}(\mu)b_\lambda(q,t)\frac{\mathfrak{h}_\mu}{z_\mu}\int_{\mathbb{T}^{N}} \prod_{i=1}^N\frac{\mathrm{d}x_i}{2\pi \mathrm{i}x_i}\omega(\x;q,t)\powsym_\mu(\x)P_\lambda(\x)\powsym_\mu(\x^{-1})P_\lambda(\x^{-1})\\
    &=\chi''_N\sum_{|\mu|\leq k}\sum_{\substack{|\nu|=|\mu|+|\lambda|\\ \ell(\nu)\leq N}}u^{|\lambda|} \mathfrak{c}_{k-|\mu|}(\mu)b_\lambda(q,t)\frac{\mathfrak{h}_\mu}{z_\mu}\left[\mathfrak{P}_{\mu,\lambda}^\nu(q,t)\right]^2\N_{\nu,N}\,.
    \end{split}
\end{align}
Here $\N_{\lambda,N}$ and $b_\lambda$ denote the normalization constant and the structure constant for ordinary Macdonald polynomials, whose explicit expressions are given in \eqref{blambdaandNormalconst}. The coefficients $\mathfrak{P}_{\mu,\lambda}^\nu(q,t)$ arise from the Pieri-type expansion of the product of a power sum symmetric function and a Macdonald polynomial:
\begin{align}
    \label{PRofPM}
    \powsym_\mu(\x)P_\lambda(\x;q,t)=\sum_{\nu}\mathfrak{P}_{\mu,\lambda}^\nu(q,t)P_\nu(\x;q,t)\,.
\end{align}
At zeroth order in $p$, the index reduces to the deformed Schur index studied in\cite{hatsuda2025deformedschurindicesmacdonald}:
\begin{align}
    \label{IN,0}
    I_{N,0}(t, u; q)=\frac{(q;q)_\infty^{N}}{(t;q)_\infty^N}\sum_{\ell(\lambda) \leq N}u^{|\lambda|} b_\lambda \N_{\lambda,N}\,.
\end{align}
For the first order correction, we need the Pieri formula for the elementary symmetric function $e_1(\mathbf{x})=\powsym_1(\mathbf{x})$. In terms of the standard Pieri rules:
\begin{align}
    \label{PRofEM}
    e_r(\x)P_\lambda(\x;q,t)=\sum_{\mu \in V_N^r(\lambda)} \psi_{\mu/\lambda}'(q,t) P_{\mu}(\x;q,t)\,,
\end{align}
where the coefficients $\psi_{\mu/\lambda}'(q,t)$ are given by
\begin{align}
\label{psi'}
\begin{split}
\psi_{\mu/\lambda}'(q,t)&=\psi_{\mu'/\lambda'}(t,q)\,,\\
\psi_{\mu/\lambda}(q,t)&=\prod_{1\leq i \leq j \leq \ell(\lambda)} \frac{(t^{j-i+1}q^{\lambda_i-\lambda_j};q)_{\mu_i-\lambda_i}(t^{j-i}q^{\lambda_i-\mu_{j+1}+1};q)_{\mu_i-\lambda_i}}{(t^{j-i}q^{\lambda_i-\lambda_j+1};q)_{\mu_i-\lambda_i}(t^{j-i+1}q^{\lambda_i-\mu_{j+1}};q)_{\mu_i-\lambda_i}}\,,
\end{split}
\end{align}
with $\lambda'$ denoting the conjugate partition of $\lambda$ and 
\begin{align}
V_{n}^r(\lambda)&=\{ \mu \vdash |\lambda|+r \; | \; \ell(\mu) \leq n \;\; \text{and} \;\; \mu/\lambda\;\; \text{is a vertical strip} \}\,.
\end{align}
Using these ingredients, the $p^1$ coefficient of the superconformal index becomes
\begin{align}
    \label{IN,1}
    \begin{split}
    I_{N,1}(t, u; q)=\frac{(1-t)(q;q)_\infty^{N}}{(1-q)(t;q)_\infty^N}\sum_{\ell(\lambda) \leq N}\sum_{\substack{\mu=A(\lambda)+\lambda \\
    \ell(\mu)\leq N}} \left[u^{|\lambda|-1}\frac{q}{t}-u^{|\lambda|}(1+\frac{q}{t})+u^{|\lambda|+1}\right] b_\lambda\psi_{\mu/\lambda}'^2\N_{\mu,N}\,,
    \end{split}
\end{align}
where $\mathrm{A}(\lambda)$ denotes the set of boxes that can be added to the Young diagram of $\lambda$ while keeping the resulting diagram a valid Young diagram with length not exceeding $N$. In other words, the sum runs over partitions $\mu$ obtained from $\lambda$ by adding a single box.

At order $p^2$, the expression is more involved. Collecting the terms from the expansion of $\Phi$ and applying the appropriate Pieri rules, we obtain
\begin{align}
    \label{I_N,2}
    I_{N,2}&=\chi''_N\sum_{\ell(\lambda) \leq N}b_\lambda \bigg[u^{|\lambda|-2}\bigg(\scalebox{1.0}{$\frac{q^2(1-t^{-1})^2}{2(1-q)^2}$}\sum_{\substack{\nu=\mu+\mathrm{A}(\mu) \\
    \ell(\nu)\leq N}}\sum_{\substack{\mu=\lambda+\mathrm{A}(\lambda) \\
    \ell(\mu)\leq N}}\psi_{\nu/\mu}'^2\psi_{\mu/\lambda}'^2\N_{\nu,N}+\scalebox{1.0}{$\frac{q^2(t^{-2}-1)}{2(1-q^2)}$}\sum_{\mu}(\mathfrak{P}_{[2],\lambda}^{\mu})^2\N_{\mu,N}\bigg)\notag\\
    +&u^{|\lambda|-1}\bigg(\frac{-(\frac{q}{t}+\frac{q^2}{t^2})(1-t)^2}{(1-q)^2}\sum_{\substack{\nu=\mu+\mathrm{A}(\mu) \\
    \ell(\nu)\leq N}}\sum_{\substack{\mu=\lambda+\mathrm{A}(\lambda) \\
    \ell(\mu)\leq N}}\psi_{\nu/\mu}'^2\psi_{\mu/\lambda}'^2\N_{\nu,N}+\frac{\frac{q}{t}(1-t)}{1-q}\sum_{\substack{\mu=\lambda+\mathrm{A}(\lambda) \\
    \ell(\mu)\leq N}}\psi_{\mu/\lambda}'^2(q,t)\N_{\mu,N}\bigg)\notag\\
    +&u^{|\lambda|}\bigg(\frac{(1-t)^2(1+(\frac{q}{t})^2+\frac{4q}{t})}{2(1-q)^2}\sum_{\substack{\nu=\mu+\mathrm{A}(\mu) \\
    \ell(\nu)\leq N}}\sum_{\substack{\mu=\lambda+\mathrm{A}(\lambda) \\
    \ell(\mu)\leq N}}\psi_{\nu/\mu}'^2\psi_{\mu/\lambda}'^2\N_{\nu,N}\notag\\
    &\quad\quad+\frac{(t-1)(1+\frac{q}{t})}{1-q}\sum_{\substack{\mu=\lambda+\mathrm{A}(\lambda) \\
    \ell(\mu)\leq N}}\psi_{\mu/\lambda}'^2(q,t)\N_{\mu,N}+\frac{(t^2-1)(1+\frac{q^2}{t^2})}{2(1-q^2)}\sum_{\mu}(\mathfrak{P}_{[2],\lambda}^{\mu})^2\N_{\mu,N}\bigg)\notag\\
    +&u^{|\lambda|+1}\bigg(\frac{-(\frac{q}{t}+1)(1-t)^2}{(1-q)^2}\sum_{\substack{\nu=\mu+\mathrm{A}(\mu) \\
    \ell(\nu)\leq N}}\sum_{\substack{\mu=\lambda+\mathrm{A}(\lambda) \\
    \ell(\mu)\leq N}}\psi_{\nu/\mu}'^2\psi_{\mu/\lambda}'^2\N_{\nu,N}+\frac{1-t}{1-q}\sum_{\substack{\mu=\lambda+\mathrm{A}(\lambda) \\
    \ell(\mu)\leq N}}\psi_{\mu/\lambda}'^2(q,t)\N_{\mu,N}\bigg)\notag\\
    +&u^{|\lambda|+2}\bigg(\frac{(1-t)^2}{2(1-q)^2}\sum_{\substack{\nu=\mu+\mathrm{A}(\mu) \\
    \ell(\nu)\leq N}}\sum_{\substack{\mu=\lambda+\mathrm{A}(\lambda) \\
    \ell(\mu)\leq N}}\psi_{\nu/\mu}'^2\psi_{\mu/\lambda}'^2\N_{\nu,N}+
    \frac{1-t^2}{2(1-q^2)}\sum_{\mu}(\mathfrak{P}_{[2],\lambda}^{\mu})^2\N_{\mu,N}\bigg)\bigg]\,.
\end{align}
In this expression, the sums over $\mu$ in the terms involving $\mathfrak{P}_{[2],\lambda}^{\mu}$ run over all partitions $\ell(\mu)\leq N$ such that the Pieri coefficient is nonvanishing. For completeness, we present examples of Pieri coefficients $\mathfrak{P}_{[2],\lambda}^{\mu}$ for the product $\powsym_2(\mathbf{x})P_\lambda(\mathbf{x})$
\begin{align}
    \begin{split}\label{PRCofPM}
    &\mathfrak{P}^{[2]}_{[2],[0]}=1\,,\quad \mathfrak{P}^{[1,1]}_{[2],[0]}=\frac{(1+q)(1-t)}{qt-1}\,,\quad \mathfrak{P}^{[3]}_{[2],[1]}=1,\quad \mathfrak{P}^{[2,1]}_{[2],[1]}=\frac{(q-t)(q+1)}{q^2t-1}\,,\\
    &\mathfrak{P}^{[1,1,1]}_{[2],[1]}=\frac{(q^2-1)(t^3-1)}{(qt-1)(1-qt^2)}\,,\quad \mathfrak{P}^{[4]}_{[2],[2]}=1\,,\quad \mathfrak{P}^{[3,1]}_{[2],[2]}=\frac{(qt-q^2)(1+q)(1-t)}{(q^3t-1)(qt-1)}\,,\\
    &\mathfrak{P}^{[2,2]}_{[2],[2]}=\frac{(q-1)^2 (q+1) (t+1) (q t^2-1)}{(q t-1)^2 (q^2 t-1)}\,,\quad\mathfrak{P}^{[2,1,1]}_{[2],[2]}=\frac{(1-q) (q+1)^2 (1-t) (1-q t^3)}{(q t-1)^3 (q t+1)}\,,\\
    &\mathfrak{P}^{[1,1,1,1]}_{[2],[2]}=0\,,\quad\mathfrak{P}^{[3,1]}_{[2],[1,1]}=1\,,\quad\mathfrak{P}^{[2,2]}_{[2],[2]}=\frac{(1+q)(1-t)}{qt-1}\,,\quad\mathfrak{P}^{[2,1,1]}_{[2],[1,1]}=\frac{(q-1) (q+1) t (q-t)}{(q t-1)^2 (q t+1)}\,,\\
    &\mathfrak{P}^{[1,1,1,1]}_{[2],[1,1]}=\frac{(1-q) (q+1) (t-1) (t^2+1) (t^2+t+1)}{(q t^2-1) (q t^3-1)}\,.
    \end{split}
\end{align}

\subsection{A deformed 1/2 BPS limit $q=t$}
We now consider the limit $q=t$. Further setting $p=0$, this would become the simple 1/2 BPS index.  So we call this case a $p$-deformed 1/2 BPS limit. In this limit, the elliptic gamma function simplifies using the identity
\begin{align}
    \Gamma(q;p,q)=\prod_{m,n=0}^\infty\frac{1-p^{m+1}q^n}{1-p^{m}q^{n+1}}=\frac{(p;p)_\infty}{(q;q)_\infty}\,,
\end{align}
Consequently, the integral representation \eqref{U(N)SCI3} reduces, for $N\geq 2$, to
\begin{align}
\label{q=t I_N}
I_N^{q=t}=\frac{1}{N!}\frac{(p;p)^{2N}_\infty}{\theta^N(u;p)}\int_{\mathbb{T}^{N}} 
\prod_{1 \leq i < j \leq N} \frac{\theta((z_i/z_j)^{\pm 1}; p)}{\theta( u(z_i/z_j)^{\pm 1}; p)} 
\prod_{j=1}^{N} 
\frac{dz_j}{2\pi i z_j}\,.
\end{align}
In this special case, the quantities $\mathfrak{h}_i$ simplify to $\mathfrak{h}_i=u^i+u^{-i}-2$, and the index becomes independent of $q$ and $t$. Moreover, the structure constants appearing in the expansion reduce to particularly simple forms:
\begin{align}
    \label{q=t structureconst}
    \psi'_{\lambda/\mu}(q,q)=1\,,\quad b_\lambda(q,q)=1\,,\quad \N_{\lambda,N}(q,q)=1\,,
\end{align}
in this limit, the coefficients listed in \eqref{PRCofPM} become $\pm 1$, while for general $\lambda,\mu,\nu$, the coefficients $\mathfrak{P}_{\lambda,\mu}^\nu(q,q)$ are integers. So in this limit, the preceding calculations simplify considerably. At zeroth order in $p$, the index reduces to the generating function of partitions with at most $N$ parts:
\begin{align}
    \label{q=t I_N,0}
    I_{N,0}(u)=\frac{1}{(u;u)_N}=\sum_{\ell(\lambda)\leq N}u^{|\lambda|}=\sum_{n=0}^\infty p_{n,N}u^n\,,
\end{align}
where $p_{n,N}$ denotes the number of partitions of $n$ with length at most $N$. This is the well known 1/2 BPS index. For the first order correction, we obtain

\begin{align}
    \label{q=t I_N,1}
    I^{q=t}_{N,1}&=\sum_{\ell(\lambda)\leq N}|\mathrm{A}(\lambda)|(u^{|\lambda|+1}+u^{|\lambda|-1}-2u^{|\lambda|})\notag\\
    &=u^{-1}+\sum_{n=1}^\infty u^n\bigg(\sum_{\substack{|\lambda|=n+1 \\
    \ell(\lambda)\leq N}}+\sum_{\substack{|\lambda|=n-1 \\
    \ell(\lambda)\leq N}}-2\sum_{\substack{|\lambda|=n \\
    \ell(\lambda)\leq N}}\bigg)|\mathrm{A}(\lambda)|\\
    &=u^{-1}+\bigg(\mathrm{A}_N(1)-2\mathrm{A}_N(0)\bigg)+\sum_{n=1}^\infty\bigg(\mathrm{A}_N(n+1)+\mathrm{A}_N(n-1)-2\mathrm{A}_N(n)\bigg) u^n\,,
    \notag
\end{align}
where $\mathrm{A}(\lambda)$ denotes the set of addable boxes of the Young diagram $\lambda$ (subject to the length constraint $\ell(\lambda)\leq N$), and $|\mathrm{A}(\lambda)|$ is its cardinality. We have introduced the notation $\mathrm{A}N(n):=\sum{\substack{|\lambda|=n \ \ell(\lambda)\leq N}}|\mathrm{A}(\lambda)|$. For $N\geq 2$, one finds $\mathrm{A}_N(1)=2$ and $\mathrm{A}_N(0)=1$, so that $\mathrm{A}_N(1)-2\mathrm{A}_N(0)=0$. Explicit values of $\mathrm{A}_N(n)$ for small $N$ are tabulated below:
\begin{align}
     \label{A_N(n)}
     &\scalebox{0.9}{$\A_2(0)=1,\  \A_2(1)=2,\ \A_2(2)=3,\ \A_2(3)=4,\ \A_2(4)=5,\ \A_2(5)=6,\ \A_2(6)=7,\ \A_2(7)=8,\ \A_2(8)=9$},\notag\\
     &\scalebox{0.89}{$\A_3(0)=1,\  \A_3(1)=2,\ \A_3(2)=4,\ \A_3(3)=6,\ \A_3(4)=9,\ \A_3(5)=12,\A_3(6)=16,\A_3(7)=20,\A_3(8)=25$},\notag\\
     &\scalebox{0.89}{$\A_4(0)=1,\  \A_4(1)=2,\ \A_4(2)=4,\ \A_4(3)=7,\ \A_4(4)=11,\A_4(5)=16,\A_4(6)=23,\A_4(7)=30,\A_4(8)=41$},\notag\\
     &\scalebox{0.89}{$\A_5(0)=1,\  \A_5(1)=2,\ \A_5(2)=4,\ \A_5(3)=7,\ \A_5(4)=12,\A_5(5)=18,\A_5(6)=27,\A_5(7)=37,\A_5(8)=53$},\notag\\
     &\scalebox{0.89}{$\A_6(0)=1,\  \A_6(1)=2,\ \A_6(2)=4,\ \A_6(3)=7,\ \A_6(4)=12,\A_6(5)=19,\A_6(6)=29,\A_6(7)=41,\A_6(8)=60$},\notag\\
     &\scalebox{0.89}{$\A_7(0)=1,\  \A_7(1)=2,\ \A_7(2)=4,\ \A_7(3)=7,\ \A_7(4)=12,\A_7(5)=19,\A_7(6)=30,\A_7(7)=43,\A_7(8)=64$},\notag\\
     &\scalebox{0.89}{$\A_8(0)=1,\  \A_8(1)=2,\ \A_8(2)=4,\ \A_8(3)=7,\ \A_8(4)=12,\A_8(5)=19,\A_8(6)=30,\A_8(7)=44,\A_8(8)=66$}.
\end{align}
Using these values, we obtain the first-order corrections for $N\leq 8$ up to order $u^7$:
\begin{align}
    \label{q=t I_2,3,4;1}
    \begin{split}
    I^{q=t}_{2,1}&=u^{-1}+0+\O(u^8)\,,\\
    I^{q=t}_{3,1}&=u^{-1}+u+0u^2+u^3+0u^4+u^5+0u^6+u^7+\O(u^8)\,,\\
    I^{q=t}_{4,1}&=u^{-1}+u+u^2+u^3+u^4+2u^5+0u^6+4u^7+\O(u^8)\,,\\
    I^{q=t}_{5,1}&=u^{-1}+u+u^2+2u^3+u^4+3u^5+u^6+6u^7+\O(u^8)\,,\\
    I^{q=t}_{6,1}&=u^{-1}+u+u^2+2u^3+u^4+3u^5+2u^6+7u^7+\O(u^8)\,,\\
    I^{q=t}_{7,1}&=u^{-1}+u+u^2+2u^3+u^4+4u^5+2u^6+8u^7+\O(u^8)\,,\\
    I^{q=t}_{8,1}&=u^{-1}+u+u^2+2u^3+u^4+4u^5+3u^6+8u^7+\O(u^8)\,.
    \end{split}
\end{align}
At second order in $p$, the general expression \eqref{I_N,2} simplifies considerably in the $q=t$ limit. Using the structure constants from \eqref{q=t structureconst}, we find
\begin{align}\label{IN,2inq=tlimit}
      I_{N,2}^{q=t}&=\sum_{\ell(\lambda) \leq N} \bigg[u^{|\lambda|-2}\bigg(\frac{1}{2}\sum_{\substack{\nu=\mu+\mathrm{A}(\mu) \\
    \ell(\nu)\leq N}}\sum_{\substack{\mu=\lambda+\mathrm{A}(\lambda) \\
    \ell(\mu)\leq N}}1+\frac{1}{2}\sum_{\mu}(\mathfrak{P}_{[2],\lambda}^{\mu}(q,q))^2\bigg)\notag\\
    &+u^{|\lambda|-1}\bigg(-2\sum_{\substack{\nu=\mu+\mathrm{A}(\mu) \\
    \ell(\nu)\leq N}}\sum_{\substack{\mu=\lambda+\mathrm{A}(\lambda) \\
    \ell(\mu)\leq N}}1+\sum_{\substack{\mu=\lambda+\mathrm{A}(\lambda) \\
    \ell(\mu)\leq N}}1\bigg)\notag\\
    &+u^{|\lambda|}\bigg(3\sum_{\substack{\nu=\mu+\mathrm{A}(\mu) \\
    \ell(\nu)\leq N}}\sum_{\substack{\mu=\lambda+\mathrm{A}(\lambda) \\
    \ell(\mu)\leq N}}1-2\sum_{\substack{\mu=\lambda+\mathrm{A}(\lambda) \\
    \ell(\mu)\leq N}}1-\sum_{\mu}(\mathfrak{P}_{[2],\lambda}^{\mu}(q,q))^2\bigg)\notag\\
    &+u^{|\lambda|+1}\bigg(-2\sum_{\substack{\nu=\mu+\mathrm{A}(\mu) \\
    \ell(\nu)\leq N}}\sum_{\substack{\mu=\lambda+\mathrm{A}(\lambda) \\
    \ell(\mu)\leq N}}1+\sum_{\substack{\mu=\lambda+\mathrm{A}(\lambda) \\
    \ell(\mu)\leq N}}1\bigg)\notag\\
    &+u^{|\lambda|+2}\bigg(\frac{1}{2}\sum_{\substack{\nu=\mu+\mathrm{A}(\mu) \\
    \ell(\nu)\leq N}}\sum_{\substack{\mu=\lambda+\mathrm{A}(\lambda) \\
    \ell(\mu)\leq N}}1+
    \frac{1}{2}\sum_{\mu}(\mathfrak{P}_{[2],\lambda}^{\mu}(q,q))^2\bigg)\bigg]\,.
\end{align}
Evaluating these combinatorial sums for various $N$, we obtain the following expansions up to order $u^4$:
\begin{align}
\begin{split}
    I^{q=t}_{2,2}&=2u^{-2}-u^{-1}+3u^0-3u+4u^2-4u^3+4u^4+\mathcal{O}(u^5)\,,\\
    I^{q=t}_{3,2}&=2u^{-2}+0u^{-1}+3u^0-u+4u^2+0u^3+4u^4+\mathcal{O}(u^5)\,,\\
    I^{q=t}_{4,2}&=2u^{-2}+0u^{-1}+4u^0-u+2u^2+7u^3-3u^4+\mathcal{O}(u^5)\,,\\
    I^{q=t}_{5,2}&=2u^{-2}+0u^{-1}+4u^0+0u+6u^2+u^3+9u^4+\mathcal{O}(u^5)\,,\\
    I^{q=t}_{6,2}&=2u^{-2}+0u^{-1}+4u^0+0u+7u^2+7u^3+20u^4+\mathcal{O}(u^5)\,,\\
    I^{q=t}_{7,2}&=2u^{-2}+0u^{-1}+4u^0+0u+7u^2+2u^3+10u^4+\mathcal{O}(u^5)\,,\\
    I^{q=t}_{N\geq8,2}&=2u^{-2}+0u^{-1}+4u^0+0u+7u^2+2u^3+11u^4+\mathcal{O}(u^5)\,.\\
\end{split}
\end{align}

Before concluding this subsection, we comment on the symmetry properties of the index. The integral representation \eqref{U(N)SCI3} is invariant under the exchange $t\leftrightarrow u$, as well as under $p\leftrightarrow q$. However, these symmetries are not manifest in our perturbative expansion, which treats $p$ as a small parameter while keeping $q$ fixed. In other words, by expanding in powers of $p$ (and $u$), we obtain a power series representation that obscures the underlying symmetries of the exact index. Nevertheless, the symmetry under $p\leftrightarrow q$ can be exploited to generate alternative expansions: one can simply interchange the roles of $p$ and $q$ in \eqref{I_N,k} to obtain an expansion in powers of $q$. In particular, taking the $p\to 0$ limit yields the deformed Schur index, while the $q\to 0$ limit followed by the replacement $p\to q$ produces a $q$-power expansion of the similar deformed Schur index. In this way we may obtain many non-trivial identities generalizing those in \cite{hatsuda2025deformedschurindicesmacdonald}, which are usually difficult to directly prove.  

Other limits also yield numerous interesting consequences. In particular, when taking $p=0$ and $u = q/t$, the superconformal index reduces to the flavored Schur index. Upon further specialization to $t = q^{1/2}$, it becomes the original unflavored Schur index, which is known to possess nice modular properties. These indices have been studied extensively in the literature, see e.g.  \cite{Bourdier:2015wda, Beem:2021zvt, Pan:2021mrw, Huang:2022bry, Hatsuda:2022xdv, Du:2023kfu, Beccaria:2024szi}.

\subsection{Large $N$ limit}
Within the framework of the AdS/CFT correspondence, understanding the large $N$ limit and its finite $N$ corrections is of central importance. However, extracting the $N\to\infty$ behavior directly from the matrix integral representations \eqref{eq:SCI} or \eqref{eq:SCI-2} is not a straightforward task. In this subsection, we adopt an alternative approach based on the theory of symmetric functions\cite{macdonald1998symmetric}, which provides a systematic framework for analyzing the strict infinite $N$ regime \cite{hatsuda2025deformedschurindicesmacdonald}.

We first introduce some useful quantities from symmetric function theory:
\begin{align}
    \label{phibvarpsi}
    \varphi_{\mu/\lambda}'(q,t)&=\frac{b_{\lambda}}{b_\mu}\psi_{\mu/\lambda}'(q,t)=\varphi_{\mu'/\lambda'}(t,q)\,,\\
\varphi_{\mu/\lambda}(q,t)&=\prod_{1\leq i \leq j \leq \ell(\mu)} \frac{(t^{j-i+1}q^{\mu_i-\mu_j};q)_{\mu_j-\lambda_j}(t^{j-i}q^{\lambda_i-\mu_{j+1}+1};q)_{\mu_{j+1}-\lambda_{j+1}}}{(t^{j-i}q^{\mu_i-\mu_j+1};q)_{\mu_j-\lambda_j}(t^{j-i+1}q^{\lambda_i-\mu_{j+1}};q)_{\mu_{j+1}-\lambda_{j+1}}}\,.
\end{align}
With these definitions, the first order coefficient \eqref{IN,1} can be rewritten as
\begin{align}
    \label{I_N,1varohi}
    \begin{split}
    I_{N,1}(t, u; q)=\scalebox{1.1}{$\frac{(1-t)(q;q)_\infty^{N}}{(1-q)(t;q)_\infty^N}$}\sum_{\ell(\lambda) \leq N}\sum_{\substack{\mu=\lambda+\mathrm{A}(\lambda) \\
    \ell(\mu)\leq N}} \left[u^{|\lambda|-1}\frac{q}{t}-u^{|\lambda|}(1+\frac{q}{t})+u^{|\lambda|+1}\right] \psi_{\mu/\lambda}'\varphi_{\mu/\lambda}'b_\mu\N_{\mu,N}\,.
    \end{split}
\end{align}
After some algebraic manipulations, the product $b_\lambda \N_{\lambda,N}$ can be simplified to
\begin{equation}
\begin{aligned}
b_\lambda \N_{\lambda,N}=\frac{(t;q)_\infty^N}{(q;q)_\infty^{N-1}(t;t)_{N}(t^Nq;q)_\infty}\prod_{i=1}^{\ell(\lambda)} \frac{(t^{N-i+1};q)_{\lambda_i}}{(t^{N-i}q;q)_{\lambda_i}}\,,
\end{aligned}
\label{eq:bN}
\end{equation}
and one finds
\begin{equation}
\begin{aligned}
\lim_{N \to \infty} \prod_{i=1}^{\ell(\lambda)} \frac{(t^{N-i+1};q)_{\lambda_i}}{(t^{N-i}q;q)_{\lambda_i}}
=1\,.
\end{aligned}
\end{equation}
Consequently, the deformed Schur index in the infinite-$N$ limit reduces to
\begin{align}
    \label{inftydeformedschur}
    I_\infty(t,u;0,q)=\lim_{N\to\infty}\bigg[\frac{(q;q)_\infty}{(t;t)_{N}(t^Nq;q)_\infty}\sum_{\ell(\lambda) \leq N} u^{|\lambda|}\prod_{i=1}^{\ell(\lambda)} \frac{(t^{N-i+1};q)_{\lambda_i}}{(t^{N-i}q;q)_{\lambda_i}}\bigg]=\frac{(q;q)_\infty}{(t;t)_\infty(u;u)_\infty}\,,
\end{align}
and
\begin{align}
    \label{I_infty,1}
    I_{\infty,1}(u,t;q)=\frac{(q^2;q)_\infty}{(t^2;t)_\infty}\sum_{\lambda}\sum_{\mu=\lambda+\mathrm{A}(\lambda)} \left[u^{|\lambda|-1}\frac{q}{t}-u^{|\lambda|}(1+\frac{q}{t})+u^{|\lambda|+1}\right] \psi_{\mu/\lambda}'\varphi_{\mu/\lambda}'(q,t)\,.
\end{align}

To proceed further, we define an expectation value and an inner product with respect to the non-elliptic Macdonald weight function, following \cite{hatsuda2025deformedschurindicesmacdonald}:
\begin{align}
\vev{A(\x)}_N'&=\frac{1}{N!} \oint_{\mathbb{T}^N} \prod_{i=1}^N \frac{dx_i}{2\pi \mathrm{i} x_i} \omega(\x) A(\x)\,,\\
\langle f, g \rangle_N'&=\frac{1}{N!} \oint_{\mathbb{T}^N} \prod_{i=1}^N \frac{dx_i}{2\pi \mathrm{i} x_i} \omega(\x) f(\x)g(\x^{-1})=\vev{f(\x)g(\x^{-1})}_N'\,,
\label{eq:inner-N}
\end{align}
where $\omega(\x)$ is the ordinary Macdonald weight. In this notation, the superconformal index and the deformed Schur index (the $p\to0$ limit) take the form
\begin{align}
I_N(t,u;p,q)&=\frac{(q;q)_\infty^N}{(t,q)_\infty^N} \biggl\langle \prod_{i,j=1}^N \frac{(tu x_i/x_j;q)_\infty}{(ux_i/x_j;q)_\infty}\Phi(\x;p,q,t,u) 
\biggr\rangle_N'\,,\\
I_N(t,0;0,q)&=\frac{(q;q)_\infty^N}{(t,q)_\infty^N}\vev{1}_N'\,.
\end{align}
We now introduce the infinite $N$ counterparts by normalizing with the vacuum expectation value:
\begin{align}
\vev{A(x)}_\infty=\lim_{N \to \infty} \frac{\vev{A(x)}_N'}{\vev{1}_N'}\,,\qquad
\langle f, g \rangle_\infty=\lim_{N \to \infty} \frac{\langle f, g \rangle_N'}{\langle 1, 1 \rangle_N'}\,.
\end{align}
In these expressions, the functions on the left hand sides depend on an infinite number of variables. The inner product $\langle f, g \rangle_\infty$ enjoys a remarkably simple property: the power sum symmetric functions become orthogonal,
\begin{equation}\label{infiniteNorthoofpowersum}
\langle p_\lambda, p_\mu \rangle_\infty=\delta_{\lambda, \mu}z_\lambda(q,t)\,,
\end{equation}
where
\begin{equation}
\begin{aligned}
z_\lambda(q,t)=z_\lambda \prod_{i=1}^{\ell(\lambda)} \frac{1-q^{\lambda_i}}{1-t^{\lambda_i}}\,.
\end{aligned}
\end{equation}
This orthogonality does not hold for finite $N$.

Consider ratio
\begin{align}
    \label{ratio}
    \frac{I_\infty(t,u;p,q)}{I_\infty(t,0;0,q)}=\lim_{N\to \infty}\frac{I_N(t,u;p,q)}{I_N(t,0;0,q)}=\biggl\langle \prod_{i,j=1}^N \frac{(tu x_i/x_j;q)_\infty}{(ux_i/x_j;q)_\infty}\Phi(\x;p,q,t,u) 
\biggr\rangle_\infty\,.
\end{align}
Using the Cauchy type identity for the Macdonald kernel,
\begin{equation}
\begin{aligned}
\prod_{i,j=1}^\infty \frac{(tu x_i/x_j;q)_\infty}{(ux_i/x_j;q)_\infty} &=\exp\biggl( \sum_{n=1}^\infty \frac{u^n}{n} \frac{1-t^n}{1-q^n}\powsym_n(\x)\powsym_n(\x^{-1}) \biggr) \\
&=\sum_{\lambda} \frac{u^{|\lambda|}}{z_\lambda(q,t)} \powsym_\lambda(\x)\powsym_\lambda(\x^{-1})\,,
\end{aligned}
\end{equation}
together with the expansion \eqref{pexph_i} of $\Phi(\x;p,q,t,u)$, we obtain an expression valid at finite $N$:
\begin{align}\label{I_N}
    \begin{split}
        I_N(t,u;p,q)&=\frac{(q;q)_\infty^N}{(t,q)_\infty^N} \biggl\langle\sum_{\lambda} \sum_{\mu}\sum_{n=0}^\infty\mathfrak{c}_n(\mu)p^{|\mu|+n}\frac{\mathfrak{h}_\mu}{z_\mu}\frac{u^{|\lambda|}}{z_\lambda(q,t)} \powsym_\lambda(\x)\powsym_\mu(\x)\powsym_\lambda(\x^{-1}) \powsym_\mu(\x^{-1})\biggr\rangle_N'\\
        &=\frac{(q;q)_\infty^N}{(t,q)_\infty^N}\sum_{\lambda} \sum_{\mu}\sum_{n=0}^\infty\mathfrak{c}_n(\mu)p^{|\mu|+n}\frac{\mathfrak{h}_\mu}{z_\mu}\frac{u^{|\lambda|}}{z_\lambda(q,t)}\biggl\langle\powsym_{\lambda\cup\mu}(\x)\powsym_{\lambda\cup\mu}(\x^{-1})\biggr\rangle_N'\,.
    \end{split}
\end{align}
For partitions written in frequency notation, $\lambda=(1^{k_1}2^{k_2}\cdots)$ and $\mu=(1^{m_1}2^{m_2}\cdots)$, the union $\lambda\cup\mu$ is simply $(1^{k_1+m_1}2^{k_2+m_2}\cdots)$. Taking the limit $N\to\infty$ in the ratio \eqref{ratio} yields
\begin{align}\label{ratioinfN}
    \frac{I_{\infty}(t,u;p,q)}{I_{\infty}(t,0;0,q)}=\sum_{\lambda} \sum_{\mu}\sum_{n=0}^\infty\mathfrak{c}_n(\mu)p^{|\mu|+n}\frac{\mathfrak{h}_\mu}{z_\mu}\frac{u^{|\lambda|}}{z_\lambda(q,t)}z_{\lambda\cup\mu}(q,t)\,.
\end{align}
To simplify this expression, we employ the identities
\begin{align}
\begin{split}
z_{\lambda \cup \mu}(q,t)&=z_\lambda(q,t)z_\mu(q,t)\prod_{i \geq 1} \binom{k_i+m_i}{k_i}\,,\\
z_{\mu}(q,t)\sum_{\lambda}\prod_{i\geq1}u^{ik_i}\binom{k_i+m_i}{k_i}=z_{\mu}&(q,t)\prod_{i\geq1}\frac{1}{(1-u^i)^{m_i+1}}=\frac{z_{\mu}}{(u;u)_\infty}\prod_{i\geq1}^{\ell(\mu)}\frac{1-q^{\mu_i}}{(1-t^{\mu_i})(1-u^{\mu_i})}\,,
\end{split}
\end{align}
substituting these into \eqref{ratioinfN} gives
\begin{align}
    \label{I_infty}
    \begin{split}
     \frac{I_{\infty}(t,u;p,q)}{I_{\infty}(t,0;0,q)}
     &=\frac{1}{(u;u)_\infty} \sum_{\mu}\sum_{n=0}^\infty\mathfrak{c}_n(\mu)p^{|\mu|+n}\mathfrak{h}_\mu\prod_{i\geq1}^{\ell(\mu)}\frac{1-q^{\mu_i}}{(1-t^{\mu_i})(1-u^{\mu_i})}\\
     &=\frac{1}{(u;u)_\infty} \sum_{\mu}\sum_{n=0}^\infty\mathfrak{c}_n(\mu)p^{|\mu|+n}\prod_{i\geq1}^{\ell(\mu)}\bigg[\left(\frac{q}{tu}\right)^{\mu_i}-1\bigg]\\
     &=\frac{1}{(u;u)_\infty}\sum_{\mu}\prod_{i\geq1}^{\ell(\mu)}\frac{(pq/tu)^{\mu_i}-p^{\mu_i}}{1-p^{\mu_i}}\,,
     \end{split}
\end{align}
recall that $I_{\infty}(t,0;0,q)=\frac{(q;q)_\infty}{(t;t)_\infty}$, so we finally obtain
\begin{align}
    \label{I_inftyresult}
    I_{\infty}(t,u;p,q)=\frac{(q;q)_\infty}{(t;t)_\infty(u;u)_\infty}\sum_{\mu}\prod_{i\geq1}^{\ell(\mu)}\frac{(pq/tu)^{\mu_i}-p^{\mu_i}}{1-p^{\mu_i}}\,,
\end{align}

To recover the famous closed form of the large $N$ superconformal index, we convert the sum over partitions $\mu$ into a sum over their frequency representations $\mu=(1^{k_1},2^{k_2},\cdots)$. The sum can be rewritten as
\begin{align}
    \sum_{k_1,k_2,\cdots=0}^\infty\prod_{i=1}^\infty\bigg(\frac{(\frac{pq}{tu})^i-p^i}{1-p^i}\bigg)^{k_i}=\prod_{i=1}^\infty\sum_{k_i=0}^\infty\bigg(\frac{(\frac{pq}{tu})^i-p^i}{1-p^i}\bigg)^{k_i}\,.
\end{align}
Assuming convergence, each geometric series can be summed, yielding
\begin{align}
    \prod_{i=1}^\infty\sum_{k_i=0}^\infty\bigg(\frac{(\frac{pq}{tu})^i-p^i}{1-p^i}\bigg)^{k_i}=\prod_{i=1}^\infty\frac{1}{1-\frac{(\frac{pq}{tu})^i-p^i}{1-p^i}}=\prod_{i=1}^\infty\frac{1-p^i}{1-(\frac{pq}{tu})^i}=\frac{(p;p)_\infty}{(\frac{pq}{tu};\frac{pq}{tu})_\infty}\,.
\end{align}
Thus, based on symmetric function methods, we have checked the exact expression for the $\mathcal{N}=4$ $U(N)$ SYM superconformal index in the large-$N$ limit
\begin{align}
    I_{\infty}(t,u;p,q)=\frac{(q;q)_\infty(p;p)_\infty}{(t;t)_\infty(u;u)_\infty(\frac{pq}{tu};\frac{pq}{tu})_\infty}\,,
\end{align}
which was first derived in \cite{kinney_2007}.

\section{Computing SCI of $\mathcal{N}=4$ SYM via solving eRS model}\label{sec5}
In this section, we develop a perturbative method for computing the superconformal index by solving the elliptic Ruijsenaars Schneider model. The central idea is to determine the $p$ expansion of the elliptic normalization constants $\N_\lambda(p,q,t)$ (up to an overall factor) and the structure constants $B_\lambda(p,q,t)$. Once these quantities are known, the superconformal index can be readily evaluated using the expression \eqref{SCISUMLAM}. Although we focus primarily on the $N=2$ case in this paper, the procedure described below is also applicable to finite $N$ more generally.

Recall that the elliptic Macdonald polynomials admit an expansion in terms of generalized monomial symmetric functions of the form
\begin{align}
    \label{empexpinmsf}
    \p_\lambda(\x;p,q,t)=\sum_{k=0}^{\infty}p^k\sum_{\mu\leq \lambda+k\phi}C^k_{\lambda\mu}(q,t)m_\mu(\x)\,.
\end{align}
For $N=2$, this expansion simplifies considerably. Writing $\widehat{\lambda}=\lambda_2-\lambda_1$, one finds
\begin{align}\label{empN=2}
    \p_\lambda(\x;p,q,t)=\sum_{k=0}^{\infty}p^k\sum_{\frac{\widehat{\lambda}}{2}\leq a\leq k, a\in \mathbb{Z}}C^k_{\lambda,\lambda+a\phi}(q,t)m_{\lambda+a\phi}(\x)\,,
\end{align}
where the sum over $a$ runs over integers satisfying the indicated bounds. Incorporating the normalization condition \eqref{normization}, we can expand these elliptic Macdonald polynomials as:
\begin{align}\label{empN=2puremonomialexpansion}
    \p_\lambda(\x;p,q,t)&=\sum_{\substack{\frac{\widehat{\lambda}}{2}\leq i<0\\ i\in \mathbb{Z}}}\sum_{k=0}^\infty p^kC_{\lambda,i,k}(q,t)m_{\lambda+i\phi}(\x)+m_\lambda(\x)+\sum_{\substack{i>0\\i\in \mathbb{Z}}}\sum_{k=i}^\infty p^kC_{\lambda,i,k}(q,t)m_{\lambda+i\phi}(\x),
\end{align}
where we have introduced the shorthand $C_{\lambda,i,k}(q,t)\equiv C^k_{\lambda,\lambda+i\phi}(q,t)$.

A useful property that simplifies calculations is the homogeneity relation
\begin{align}
    \label{x1..xNmlambda,Plambda}
    \begin{split}
    &(x_1\cdots x_N)^m \mathfrak{F}_{\lambda}(\x)=\mathfrak{F}_{\lambda+m^N}(\x)\quad\quad\quad \text{For $\mathfrak{F}_{\lambda}=m_{\lambda},P_{\lambda},\p_{\lambda}$}\,.
    \end{split}
\end{align}
In the $N=2$ case, this implies that all coefficients $C_{\lambda,i,k}(q,t)$ can be deduced from those associated with partitions of the form $[n,0]$ ($n\in\mathbb{N}$), since
\begin{align}
    C_{\lambda+(l)^2,i,k}(q,t)=C_{\lambda,i,k}(q,t)\,,
\end{align}
where $(l)^2$ denotes $[l,l]$. Thus it suffices to determine the coefficients for the one-parameter family of partitions $\lambda=[n,0]$.

To compute these coefficients, we expand the elliptic Ruijsenaars-Schneider Hamiltonian in powers of $p$. For $N=2$, the relevant factor $\frac{\theta(tx_1/x_2;p)}{\theta(x_1/x_2;p)}$ admits the series expansion
\begin{align}
\begin{split}
    &\frac{\theta(tx_1/x_2;p)}{\theta(x_1/x_2;p)}=\left(\frac{tx_1-x_2}{x_1-x_2}\right)\bigg(1-\frac{ (t-1) \left(t x_1^2-x_2^2\right)}{t x_1 x_2}p-\frac{(t-1) \left(x_1^2+x_1 x_2+x_2^2\right) \left(t x_1^2-x_2^2\right)}{t x_1^2 x_2^2}p^2 \\
    &+\scalebox{1.0}{$\frac{(t-1) \left(t^3 x_1^5 x_2-t^2 x_1^2 \left(x_1^4+x_1^3 x_2+2 x_1^2 x_2^2+2 x_1 x_2^3+x_2^4\right)+t x_2^2 \left(x_1^4+2 x_1^3 x_2+2 x_1^2 x_2^2+x_1 x_2^3+x_2^4\right)-x_1 x_2^5\right)}{t^2 x_1^3 x_2^3}$}p^3+O\left(p^4\right) \bigg)\,. 
\end{split}
\end{align}
Equipped with this expansion, one can solve the eigenvalue problem for the elliptic Ruijsenaars-Schneider model perturbatively. Writing the Hamiltonian as $\mathcal{D}_x(p)=\sum_{l=0}^\infty p^l\mathcal{D}_l$ and the eigenvalues as $\varepsilon_\lambda(p,q,t)=\sum_{l=0}^\infty p^l\varepsilon_{\lambda,l}(q,t)$, the eigenfunction equation takes the form
\begin{align}
\begin{split}
    &\sum_{l=0}^\infty p^l\D_l\left[\sum_{\substack{\frac{\widehat{\lambda}}{2}\leq i<0\\ i\in \mathbb{Z}}}\sum_{k=0}^\infty p^kC_{\lambda,i,k}(q,t)m_{\lambda+i\phi}(\x)+m_\lambda(\x)+\sum_{\substack{i>0\\i\in \mathbb{Z}}}\sum_{k=i}^\infty p^kC_{\lambda,i,k}(q,t)m_{\lambda+i\phi}(\x)\right]=\\
    &\sum_{l=0}^\infty p^l\varepsilon_{\lambda,l}(q,t)\left[\sum_{\substack{\frac{\widehat{\lambda}}{2}\leq i<0\\ i\in \mathbb{Z}}}\sum_{k=0}^\infty p^kC_{\lambda,i,k}(q,t)m_{\lambda+i\phi}(\x)+m_\lambda(\x)+\sum_{\substack{i>0\\i\in \mathbb{Z}}}\sum_{k=i}^\infty p^kC_{\lambda,i,k}(q,t)m_{\lambda+i\phi}(\x)\right]\,.
    \end{split}
\end{align}
Matching coefficients order by order in $p$ yields a recursive system that determines the coefficients $C_{\lambda,i,k}(q,t)$ and the eigenvalue corrections $\varepsilon_{\lambda,l}(q,t)$. Solving this system perturbatively provides explicit $p$ expansions of the elliptic Macdonald polynomials in terms of generalized monomial symmetric functions. For illustration, we present the results for low orders in $p$:
\begin{align}
 \p_{[0,0]}&
\begin{aligned}[t]
    &=m_{[0,0]}+p\frac{q (t-1)^2 (t+1)}{(q-1) t (q t-1)}m_{[1,-1]}\notag\\
    &+p^2\left[\frac{q \left(q^2-1\right) \left(t^2-1\right)^2 \left(q t^2-1\right)}{t (q t-1)^3 \left(q^2 t-1\right)}m_{[1,-1]}+\frac{q^2 (t-1)^2 (t+1) \left(q t^2-1\right)}{(q-1)^2 (q+1) t^2 \left(q^2 t-1\right)}m_{[2,-2]}\right]\notag\\
    &+p^3\bigg[\scalebox{1.1}{$\frac{q^2 (q+1) (t-1)^3 (t+1)^2 (q t+1) \left(q t^2-1\right)}{t^2 (q t-1)^2 \left(q^2 t-1\right) \left(q^3 t-1\right)}$}m_{[2,-2]}+ \scalebox{1.1}{$\frac{q^3 (t-1)^2 (t+1) (q t-1) (q t+1) \left(q t^2-1\right)}{(q-1)^3 (q+1) \left(q^2+q+1\right) t^3 \left(q^3 t-1\right)}$}m_{[3,-3
    ]}\notag\\
    &\quad\quad+\frac{q (t-1)^2 (t+1) \left(q t^2-1\right) A}{(q-1)^3 (q+1) t^3 (q t-1)^5 \left(q^2 t-1\right) \left(q^3 t-1\right)}m_{[1,-1]}\bigg] +\mathcal{O}(p^4)\,,
\end{aligned}\\
   \p_{[1,0]}&
\begin{aligned}[t]
    &=m_{[1,0]}+p\frac{q (t-1) \left(q t^2-1\right)}{(q-1) t \left(q^2 t-1\right)}m_{[2,-1]}+p^2\bigg[\frac{q (t-1)^2 (q t-1) (q t+1) B}{(q-1)^2 (q+1) t^2 \left(q^2 t-1\right)^3 \left(q^3 t-1\right)}m_{[2,-1]}\notag\\
    &\quad\quad\quad+\frac{q^2 (t-1) (q t-1)^2 (q t+1) \left(q t^2-1\right)}{(q-1)^2 (q+1) t^2 \left(q^2 t-1\right) \left(q^3 t-1\right)}m_{[3,-2]}\bigg]\notag\\
    &+p^3\bigg[\frac{q (q+1) \left(q^2+q+1\right) (t-1) (q-t) (q t-1) (q t+1)^2 \left(q t^2-1\right) C}{t^2 \left(q^2 t-1\right)^5 \left(q^3 t-1\right) \left(q^4 t-1\right)}m_{[2,-1]}\notag\\
    &\quad\quad+\frac{q^2 (t-1) \left(q t^2-1\right) \left(q^3 t^2-1\right) D}{(q-1)^3 \left(q^2+q+1\right) t^3 \left(q^2 t-1\right)^3 \left(q^3 t-1\right) \left(q^4 t-1\right)}m_{[3,-2]}\notag\\
    &\quad\quad+\frac{q^3 (t-1) (q t-1)^2 (q t+1) \left(q t^2-1\right) \left(q^3 t^2-1\right)}{(q-1)^3 (q+1) \left(q^2+q+1\right) t^3 \left(q^3 t-1\right) \left(q^4 t-1\right)}m_{[4,-3]}\bigg]+\mathcal{O}(p^4)\,,
\end{aligned}\\
\p_{[2,0]}&
\begin{aligned}[t]
        &=m_{[2,0]}+\scalebox{1.1}{$\frac{(q+1) (t-1)}{q t-1}$}m_{[1,1]}+p\bigg[\scalebox{1.1}{$\frac{q (t-1) (q t-1) (q t+1)}{(q-1) t \left(q^3 t-1\right)}$}m_{[3,-1]}+\scalebox{1.1}{$\frac{(t-1) E}{(q-1) t (q t-1)^3 \left(q^3 t-1\right)}$}m_{[1,1]}\bigg]\notag\\
        &+p^2\bigg[\scalebox{1.1}{$\frac{(t-1) F}{(q-1)^2 t^2 (q t-1)^5 \left(q^3 t-1\right)^3 \left(q^4 t-1\right)}$}m_{[1,1]}+\scalebox{1.1}{$\frac{q (t-1) \left(q^3 t^2-1\right) G}{(q-1)^2 t^2 (q t-1) \left(q^3 t-1\right)^3 \left(q^4 t-1\right)}$}m_{[3,-1]}\notag\\
        &\quad\quad+\frac{q^2 (t-1) (q t-1)^2 (q t+1) \left(q^3 t^2-1\right)}{(q-1)^2 (q+1) t^2 \left(q^3 t-1\right) \left(q^4 t-1\right)}m_{[4,-2]}\bigg]\notag\\
        &+p^3\bigg[-\frac{(t-1) H}{(q-1)^3 t^3 (q t-1)^7 \left(q^3 t-1\right)^5 \left(q^4 t-1\right) \left(q^5 t-1\right)}m_{[1,1]}\notag\\
        &\quad\quad+\frac{q (q-t) (t-1)^2 \left(q^3 t^2-1\right) I}{(q-1)^3 (q+1) \left(q^2+q+1\right) t^3 (q t-1)^3 \left(q^3 t-1\right)^5 \left(q^4 t-1\right) \left(q^5 t-1\right)}m_{[3,-1]}\notag\\
        &\quad\quad+\frac{q^2 (q+1) (t-1)^2 (q t-1) (q t+1) \left(q^2 t-1\right) \left(q^2 t+1\right) J}{(q-1)^3 \left(q^2+q+1\right) t^3 \left(q^3 t-1\right)^3 \left(q^4 t-1\right) \left(q^5 t-1\right)}m_{[4,-2]}\notag\\
        &\quad\quad+\frac{q^3 (t-1) (q t-1)^2 (q t+1) \left(q^2 t-1\right)^2 \left(q^2 t+1\right) \left(q^3 t^2-1\right)}{(q-1)^3 (q+1) \left(q^2+q+1\right) t^3 \left(q^3 t-1\right) \left(q^4 t-1\right) \left(q^5 t-1\right)}m_{[5,-3]}\bigg]+\mathcal{O}(p^4)\,,
    \end{aligned}
\end{align}
where $A,B,C,D,E,F,G,H,I,J$ is attached at Appendix \ref{appendixA}.

As is well known, in the limit $q=t$, ordinary Macdonald polynomials reduce to Schur polynomials. This property extends naturally to the elliptic setting: in the same limit, the elliptic Macdonald polynomials become elliptic Schur polynomials, and all coefficients $C_{\lambda,i,k}(q,t)$ simplify to integers. For the $N=2$ case, we obtain the following explicit expansions up to order $p^8$:
\begin{align}\label{ellipticschurp}
        \s_{[0,0]}=&m_{[0,0]}+(p+p^2+p^3-p^5-3p^6-4p^7-4p^8)m_{[1,-1]}\notag\\
        &+(p^2+p^3+2p^4+p^5+p^6-3p^7-6p^8)m_{[2,-2]}+(p^3+p^4+2p^5+2p^6+p^7-2p^8)m_{[3,-3]}\notag\\
        &+(p^4+p^5+2p^6+2p^7+2p^8)m_{[4,-4]}+(p^5+p^6+2p^7+2p^8)m_{[5,-5]}\notag\\
        &+(p^6+p^7+2p^8)m_{[6,-6]}+(p^7+p^8)m_{[7,-7]}+p^8m_{[8,-8]}+\O(p^9)\,,\notag\\
        \s_{[1,0]}=&m_{[1,0]}+(p+p^2-p^5-p^6-p^7)m_{[2,-1]}\notag\\
        &+(p^2+p^3+p^4-2p^7-2p^8)m_{[3,-2]}+(p^3+p^4+p^5+p^6-p^8)m_{[4,-3]}\notag\\
        &+(p^4+p^5+p^6+p^7+p^8)m_{[5,-4]}+(p^5+p^6+p^7+p^8)m_{[6,-5]}\notag\\
        &+(p^6+p^7+p^8)m_{[7,-6]}+(p^7+p^8)m_{[8,-7]}+p^8m_{[9,-8]}+\O(p^9)\,,\notag\\
        \s_{[2,0]}=&(1+p-p^2-p^3+p^4+p^5+2p^6+2p^7)m_{[1,1]}+m_{[2,0]}+(p+p^2-p^4-p^5+p^7+p^8)m_{[3,-1]}\notag\\
        &+(p^2+p^3+p^4-p^5-p^6-p^7)m_{[4,-2]}+(p^3+p^4+p^5-p^7-p^8)m_{[5,-3]}\notag\\
        &+(p^4+p^5+p^6)m_{[6,-4]}+(p^5+p^6+p^7)m_{[7,-5]}\notag\\
        &+(p^6+p^7+p^8)m_{[8,-6]}+(p^7+p^8)m_{[9,-7]}+p^8m_{[10,-8]}+\O(p^9)\,.
\end{align}

To compute the normalization constants as power series in $p$, we first expand the elliptic weight function. Starting from the identity
\begin{align}
\begin{split}
     \prod_{1 \leq i\neq j \leq N}\frac{\Gamma(tx_i/x_j;p,q)}{\Gamma(x_i/x_j;p,q)}&=c_N(p,q,t)\cdot\omega(\x)\e\bigg[\sum_{n=1}^\infty\frac{\left(t^n-1\right)\left(1+\frac{q^n}{t^n}\right)p^n}{n(1-q^n)(1-p^n)}\powsym_n(\x)\powsym_n(\x^{-1})\bigg]\\
     &=c_N(p,q,t)\cdot\omega(\x)\sum_{\mu}\sum_{n=0}^\infty\mathfrak{c}_n(\mu)p^{|\mu|+n}\frac{\mathfrak{g}_\mu}{z_\mu}\powsym_\mu(\x)\powsym_\mu(\x^{-1})\,,
     \end{split}
\end{align}
where the prefactor
\begin{align}
    c_N(p,q,t)=\bigg[\frac{1}{\Gamma(t;p,q)(t;q)_\infty(p;p)_\infty}\bigg]^N\,,\quad \mathfrak{g}_{n}=\frac{(t^n-1)(1+\frac{q^n}{t^n})}{1-q^n}\,, \quad\mathfrak{g}_\mu=\prod_{i=1}^{\ell(\mu)}\mathfrak{g}_{\mu_i}\,.
\end{align}
Specializing to the $N=2$ case, the elliptic normalization constants can then be expressed as
\begin{align}\label{orthoN=2}
    \begin{split}
       \N_{\lambda,2}(p,q,t)&=c_2(p,q,t)\sum_{k_1,k_2=0}^{\infty}\sum_{\substack{\frac{\widehat{\lambda}}{2}\leq a_1\leq k_1, a_1\in \mathbb{Z}_{k_1}^*\\
       \frac{\widehat{\lambda}}{2}\leq a_2\leq k_2, a_2\in \mathbb{Z}_{k_2}^*}}\sum_{\mu}\sum_{n=0}^\infty\mathfrak{c}_n(\mu)p^{|\mu|+n+k_1+k_2}C_{\lambda,a_1,k_1}C_{\lambda,a_2,k_2}\frac{\mathfrak{g}_\mu}{z_\mu}\\
       &\times\oint_{\mathbb{T}^2} \prod_{j=1}^{2} 
       \frac{dx_j}{2\pi \mathrm{i} x_j}\omega(\x)\powsym_\mu(\x)m_{\lambda+a_1\phi}(\x)\powsym_\mu(\x^{-1})m_{\lambda+a_2\phi}(\x^{-1})\,.
    \end{split}
\end{align}
Here $\mathbb{Z}_{k}^*$ denotes $\mathbb{Z}$ when $k=0$ and $\mathbb{Z}_{\neq0}$ otherwise. Expanding the products of power sums and monomial symmetric functions in terms of Macdonald polynomials and invoking the orthogonality relation for ordinary Macdonald polynomials, we obtain
\begin{align}
    \label{ELLNORinMACNOR}
    \begin{split}
    \N_{\lambda,2}(p,q,t)&=c_2(p,q,t)\sum_{k_1,k_2=0}^{\infty}\sum_{\substack{\frac{\widehat{\lambda}}{2}\leq a_1\leq k_1, a_1\in \mathbb{Z}_{k_1}^*\\
       \frac{\widehat{\lambda}}{2}\leq a_2\leq k_2, a_2\in \mathbb{Z}_{k_2}^*}}\sum_{\mu}\sum_{n=0}^\infty\mathfrak{c}_n(\mu)p^{|\mu|+n+k_1+k_2}C_{\lambda,a_1,k_1}C_{\lambda,a_2,k_2}\frac{\mathfrak{g}_\mu}{z_\mu}\\
       &\times\sum_{\substack{\nu\\ \ell(\nu)\leq 2}}\mathfrak{M}_{\mu,\lambda+a_1\phi+l_{(a_1,a_2)}^2}^\nu\mathfrak{M}_{\mu,\lambda+a_2\phi+l_{(a_1,a_2)}^2}^\nu\N_{\nu,2}(q,t)\,,
    \end{split}
\end{align}
where the coefficients $\mathfrak{M}_{\lambda,\mu}^\nu$ are defined by the expansion
\begin{align}
    \label{mpowermacdPR}
    \powsym_{\lambda}(\x)m_{\mu}(\x)=\sum_{\nu}\mathfrak{M}_{\lambda,\mu}^\nu P_{\nu}(\x)\,,
\end{align}
and $l_{(a_1,a_2)}=\max\{a_1,a_2\}$ is the minimal integer ensuring that both $\lambda+a_1\phi+l_{(a_1,a_2)}^2$ and $\lambda+a_2\phi+l_{(a_1,a_2)}^2$ are partitions with nonnegative parts. In the $N=2$ case, the relations
\begin{align}
    \mathfrak{M}_{\lambda,\mu}^\nu =\mathfrak{M}_{\lambda,\mu+l^2}^{\nu+l^2 },\quad\quad \N_{\nu,2}(q,t)=\N_{\nu+l^2,2}(q,t)\,,
\end{align}
guarantee that the final result is independent of the choice of $l_{(a_1,a_2)}$.

Carrying out the computation to low orders in $p$, we obtain explicit expressions for the elliptic normalization constants in terms of  Macdonald normalization constants. For illustration, we present the results for several generalized partitions $\lambda\in \Lambda^2$, where $\Lambda^2=\{\lambda\in\mathbb{Z}^2\mid \lambda_1\geq \lambda_2\}$:
\begin{align}\label{N(p,q,t)}
        \N_{[0,0]}(p,q,t)&=c_2\Bigg[\N_{[0,0]}(q,t)+p\bigg[\scalebox{1.1}{$\frac{(t-1)(1+q/t)}{1-q}$}\N_{[1,0]}(q,t)-\scalebox{1.1}{$\frac{2q (q+1) (t-1)^3 (t+1)}{(q-1) t (q t-1)^2}$}\N_{[0,0]}(q,t)\bigg]\notag\\
        &\quad\quad+p^2\bigg[-\frac{(t-1) (q-t) K}{(q-1)^2 (q+1) t^2 (q t-1)^4 \left(q^2 t-1\right)^2}\N_{[0,0]}(q,t)\notag\\
        &\quad\quad\quad\quad-\scalebox{1.1}{$\frac{(t-1) (q+t)L}{(q-1)^2 t^2 (q t-1) \left(q^2 t-1\right)}$}\N_{[1,0]}(q,t)-\scalebox{1.1}{$\frac{(t-1)M}{(q-1)^2 (q+1) t^2 (q t-1)^2}$}\N_{[2,0]}(q,t)\bigg]\notag\\
        &\quad\quad+\O(p^3)\Bigg]\,,\\
       \N_{[1,-1]}(p,q,t)&=c_2\Bigg[\N_{[1,-1]}(q,t)+p\bigg[-\frac{(q-1) (q+1)^2 (t-1) (q+t) \left(q t^2-1\right)^2}{t (q t-1)^2 \left(q^2 t-1\right)^2}\N_{[1,0]}(q,t)\notag\\
       &\quad\quad\quad-\scalebox{1.1}{$\frac{2 q (q+1) \left(q^2+1\right) (t-1)^2 (q t-1) (q t+1)}{(q-1) t \left(q^3 t-1\right)^2}$}\N_{[2,0]}(q,t)-\scalebox{1.1}{$\frac{(t-1) (q+t)}{(q-1) t}$}\N_{[3,0]}(q,t)\bigg]\notag\\
       &\quad\quad+\O(p^2)\Bigg]\,,\\
       \N_{[2,-2]}(p,q,t)&=c_2\bigg[\N_{[2,-2]}(q,t)+\O(p^1)\bigg]\,,
\end{align}
where
\begin{align*}
       K&=3 q^{10} t^6+q^{10} t^4+q^9 t^8+3 q^9 t^7-7 q^9 t^6-13 q^9 t^5+2 q^9 t^4-2 q^9 t^3+8 q^8 t^8+q^8 t^7-21 q^8 t^6\\
       &+12 q^8 t^5+25 q^8 t^4-5 q^8 t^3+3 q^7 t^8-29 q^7 t^7-17 q^7 t^6+56 q^7 t^5+7 q^7 t^4-27 q^7 t^3+7 q^7 t^2\\
       &+4 q^6 t^8
       -23 q^6 t^7+25 q^6 t^6+66 q^6 t^5-70 q^6 t^4-49 q^6 t^3+25 q^6 t^2-2 q^6 t-14 q^5 t^7+49 q^5 t^6\\
       &+26 q^5 t^5-90 q^5 t^4+26 q^5 t^3+49 q^5 t^2-14 q^5 t-2 q^4 t^7+25 q^4 t^6-49 q^4 t^5-70 q^4 t^4+66 q^4 t^3\\
       &+25 q^4 t^2-23 q^4 t+4 q^4+7 q^3 t^6-27 q^3 t^5+7 q^3 t^4+56 q^3 t^3-17 q^3 t^2-29 q^3 t+3 q^3-5 q^2 t^5\\
       &+25 q^2 t^4+12 q^2 t^3-21 q^2 t^2+q^2 t+8 q^2-2 q t^5+2 q t^4-13 q t^3-7 q t^2+3 q t+q+t^4+3 t^2\,,\\
       L&=q^4 t^3+q^3 t^3-3 q^3 t^2-2 q^3 t+2 q^3-2 q^2 t^4+4 q^2 t^3-4 q^2 t+2 q^2-2 q t^4+2 q t^3+3 q t^2-q t-t,\\
       M&=q^5 t^2-q^4 t^4-2 q^4 t-q^3 t^5+q^3 t^4+5 q^3 t^3-2 q^3 t^2-q^3 t+2 q^3-2 q^2 t^5+q^2 t^4+2 q^2 t^3\\
       &-5 q^2 t^2-q^2 t+q^2+2 q t^4+q t-t^3\,.
\end{align*}
For the purpose of computing the superconformal index at $u^0$ up to order $p^2$, it suffices to retain $\N_{[1,-1]}$ to order $p^1$ and $\N_{[2,-2]}$ to order $p^0$, as will become clear from the structure constants $B_\lambda(p,q,t)$ to which we now turn.

To determine $B_\lambda(p,q,t)$, we first express the elliptic kernel function \eqref{ellkerfinexp} in a form suitable for expansion. Using the identity
\begin{align}
   \mathrm{log}\frac{\Gamma(ux_iy_j; p, q)}{\Gamma(tux_iy_j; p, q)}=\sum_{k=1}^\infty\frac{1-t^k}{k(1-p^k)(1-q^k)}\bigg[u^k(x_iy_j)^k+(\frac{pq}{tu})(x_iy_j)^{-k}\bigg]\,,
\end{align}
we obtain
\begin{align}
\begin{split}
     \mathrm{log}\prod_{1\leq i,j\leq N}\frac{\Gamma(ux_iy_j; p, q)}{\Gamma(tux_iy_j; p, q)}&=\sum_{k=1}^\infty\frac{1-t^k}{k(1-p^k)(1-q^k)}\bigg[u^k\sum_{1\leq i,j\leq N}(x_iy_j)^k+(\frac{pq}{tu})\sum_{1\leq i,j\leq N}(x_iy_j)^{-k}\bigg]\\
     &=\sum_{k=1}^\infty\frac{1-t^k}{k(1-p^k)(1-q^k)}\bigg[u^k\powsym_k(\x)\powsym_k(\y)+(\frac{pq}{tu})\powsym_k(\x^{-1})\powsym_k(\y^{-1})\bigg]\,.
\end{split}
\end{align}
Employing the Cauchy type expansion
\begin{align}
    \exp\left(\sum_{k=1}^\infty\frac{a_k}{k}\powsym_k(\x)\powsym_k(\y)\right)=\sum_{\lambda}\frac{1}{z_\lambda}a_{\lambda}\powsym_\lambda(\x)\powsym_\lambda(\y)\,,
\end{align}
the kernel function can be rewritten as
\begin{align}
    K_u(\x;\y)=\sum_{\lambda,\mu}\frac{1}{z_\lambda z_\mu}A_\lambda B_\mu\powsym_\lambda(\x)\powsym_\mu(\x^{-1})\powsym_\lambda(\y)\powsym_\mu(\y^{-1})\,,
\end{align}
where $z_\lambda$ and $z_\mu$ are defined in \eqref{defofzmupmu}, and $A_\lambda$, $B_\mu$ are constructed analogously to $g_\mu$ in \eqref{defofzmupmu} and
\begin{align}
    A_k=u^k\frac{1-t^k}{(1-q^k)(1-p^k)},\quad\quad\quad B_k=u^{-k}p^k\frac{(1-t^k)q^k}{(1-q^k)t^k(1-p^k)}\,.
\end{align}
Expanding in powers of $u$ yields
\begin{align}
    K_u(\x;\y)=\sum_{\lambda,\mu}\frac{u^{|\lambda|-|\mu|}}{z_\lambda z_\mu}p^{\sum_{i=1}^{\ell(\lambda)}\sum_{l_i=0}^\infty l_i\lambda_i+\sum_{j=1}^{\ell(\mu)}\sum_{m_j=1}^\infty m_j\mu_j}U_\lambda V_\mu\powsym_\lambda(\x)\powsym_\mu(\x^{-1})\powsym_\lambda(\y)\powsym_\mu(\y^{-1})\,,
\end{align}
with
\begin{align}
    U_k=\frac{1-t^k}{1-q^k}\,,\quad\quad V_k=\frac{(1-t^k)q^k}{(1-q^k)t^k}\,,
\end{align}
The coefficient of $u^n$ is therefore 
\begin{align}
    K^{n}_u(\x;\y)=\sum_{|\lambda|-|\mu|=n}\frac{1}{z_\lambda z_\mu}p^{\sum_{i=1}^{\ell(\lambda)}\sum_{l_i=0}^\infty l_i\lambda_i+\sum_{j=1}^{\ell(\mu)}\sum_{m_j=1}^\infty m_j\mu_j}U_\lambda V_\mu\powsym_\lambda(\x)\powsym_\mu(\x^{-1})\powsym_\lambda(\y)\powsym_\mu(\y^{-1})\,.
\end{align}
From this expression, we observe that the lowest power of $p$ accompanying a given $\mu$ is $|\mu|$. Consequently, to obtain $K^{n}_u(\mathbf{x};\mathbf{y})$ up to order $p^r$, it suffices to sum over partitions $\lambda,\mu$ satisfying $|\lambda|-|\mu|=n$ and $|\mu|\leq r$.

Using the fact that $\powsym_n(\mathbf{x})=m_n(\mathbf{x})$, and products of monomial symmetric functions can be expanded as linear combinations of monomial symmetric functions. In the $N=2$ case, the products of $m_\lambda(\x)\cdot m_\mu(\x^{-1})$ are easily computed and re-expressed in terms of generalized monomial symmetric functions. From the expansion \eqref{empN=2puremonomialexpansion}, we can solve the generalized monomial symmetric functions in terms of the elliptic Macdonald polynomials themselves. Then substituting the generalized monomial symmetric functions appear in the kernel function expansion, this leads to an expansion of the form
\begin{align}
    \label{{KernelfunctioninELLMACexpansion}}
    K_u(\x;\y)=\sum_{\lambda\in \Lambda^2}u^{|\lambda|}B_{\lambda,\mu}(p,q,t)\p_\lambda(\x;p,q,t)\p_\mu(\y;p,q,t)\,.
\end{align}
The coefficients $B_{\lambda,\mu}(p,q,t)$ are expressed in terms of the $C_{\lambda,i,k}$. For example, one finds
\begin{align}
    \label{BlaminCexpansion}
    B_{[0,0]}&=1+\scalebox{1.2}{$\frac{p \left(q (t-1)^2 C_{[1,-1],-1,0}^2+2 C_{[1,-1],-1,0} \left(t(q-1)^2 C_{[0,0],1,1}-2 q (t-1)^2\right)+4 q (t-1)^2\right)}{(q-1)^2 t}$}\notag\\
    &+\O(p^2),\\
    B_{[1,-1]}&=\frac{p q (t-1)^2}{(q-1)^2 t}+\frac{p^2}{(q-1)^4 (q+1)^2 t^2} \bigg(2 q \left(q^2-1\right)^2 (t-1)^2 t C_{[0,0],1,1} (C_{[1,-1],-1,0}-2)\notag\\
    &+q (t-1)^2 \bigg(q (q t-1)^2 C_{[2,-2],-1,0}^2+2 (q+1) C_{[2,-2],-1,0} \bigg((q-1)^2 (q+1) t C_{[1,-1],1,1}\notag\\
    &-2 q (t-1) (q t-1)\bigg)+2 (q+1) \left(q^3 t+q^2 \left(3 t^2-5 t+1\right)+q \left(t^2-5 t+3\right)+t\right)\bigg)\notag\\
    &+(q+1)^2 (q-1)^4 t^2 C_{[0,0],1,1}^2\bigg)+\O(p^3)\,.
\end{align}
Solving the elliptic Ruijsenaars Schneider model provides explicit values for the coefficients $C_{\lambda,i,k}\,$. Substituting these into the expressions for $B_{\lambda,\mu}(p,q,t)$, one verifies that $B_{\lambda,\mu}(p,q,t)=0$ for $\lambda\neq\mu$, thereby recovering the diagonal form anticipated in \eqref{ellpcauchy}. The resulting expansions for the diagonal coefficients, needed for the superconformal index at $u^0$ up to order $p^2$, are
\begin{align}\label{Blaminpexpexplicity}
        B_{[0,0]}&=1+p\frac{ q (t+1) (t-1)^2 (3 q t-q+t-3)}{(q-1) t (q t-1)^2}+p^2\frac{ q (t-1)^2  (1 + t)}{(q-1)^2 (q+1) t^2 (q t-1)^4 \left(q^2 t-1\right)^2}\notag\\
        &\times\bigg(2 t \left(-t^2+t+2\right)+2 q^9 t^4 \left(2 t^2+t-1\right)+q^8 t^2 \left(7 t^5-11 t^4-7 t^3+5 t^2-7 t+3\right)\notag\\
        &+\scalebox{0.95}{$q^7 t^2 \left(9 t^5-41 t^4+t^3+35 t^2-21 t+7\right)+q^6 t \left(6 t^6-28 t^5+58 t^4+44 t^3-41 t^2+15 t-4\right)$}\notag\\
        &-\scalebox{0.95}{$q^5 t \left(20 t^5-72 t^4+42 t^3+81 t^2-47 t+10\right)-q^4 t \left(10 t^5-47 t^4+81 t^3+42 t^2-72 t+20\right)$}\notag\\
        &+\scalebox{0.95}{$q^3 \left(-4 t^6+15 t^5-41 t^4+44 t^3+58 t^2-28 t+6\right)+q^2 \left(7 t^5-21 t^4+35 t^3+t^2-41 t+9\right)$}\notag\\
        &+q \left(3 t^5-7 t^4+5 t^3-7 t^2-11 t+7\right)\bigg)+\O(p^3)\,,\\
        B_{[1,-1]}&=p\frac{ q (t-1)^2}{(q-1)^2 t}+p^2\frac{ q (t-1)^2}{(q-1)^3 (q+1) t^2 (q t-1)^2 \left(q^3 t-1\right)^2}\bigg(2 q^{10} t^5+4 q^9 t^6-6 q^9 t^5-q^9 t^4\notag\\
        &-\scalebox{0.95}{$q^9 t^2+6 q^8 t^6-22 q^8 t^5+14 q^8 t^4+2 q^8 t^3+5 q^7 t^6-16 q^7 t^5+18 q^7 t^4-4 q^7 t^3-3 q^7 t^2+2 q^6 t^6$}\notag\\
        &-\scalebox{0.95}{$14 q^6 t^5+14 q^6 t^4+10 q^6 t^3-12 q^6 t^2+6 q^6 t+q^5 t^6-10 q^5 t^5+29 q^5 t^4-29 q^5 t^2+10 q^5 t-q^5$}\notag\\
        &-\scalebox{0.95}{$6 q^4 t^5+12 q^4 t^4-10 q^4 t^3-14 q^4 t^2+14 q^4 t-2 q^4+3 q^3 t^4+4 q^3 t^3-18 q^3 t^2+16 q^3 t-5 q^3$}\notag\\
        &-2 q^2 t^3-14 q^2 t^2+22 q^2 t-6 q^2+q t^4+q t^2+6 q t-4 q-2 t\bigg)+\O(p^3)\,,\\
        B_{[2,-2]}&=p^2\frac{ q^2 (t-1)^2 (q t-1)^2}{(q-1)^4 (q+1)^2 t^2}+\O(p^3)\,.
\end{align}

Writing the superconformal index as a power series in $u$,
\begin{align}
    \label{SCIinuexpansion}
    I_N(p,q,t,u)=\sum_{k=-\infty}^\infty u^kI^k_N(p;q,t)\,,
\end{align}
for example, we can now assemble the $u^0$ coefficient for $N=2$. Substituting the expansions of elliptic normalization constants $\N_{\lambda}(p,q,t)$ and $B_{\lambda}(p,q,t)$ given above into \eqref{SCISUMLAM} and collecting terms of order $p^0$, $p^1$, and $p^2$, we obtain
\begin{align}
    \label{I^0_N}
    \begin{split}
       I^0_2=\frac{(q;q)^2_\infty}{(t;q)^2_\infty} &\Bigg[\N_{[0,0]}(q,t)+p\bigg[\frac{ q (t-1)^2}{(q-1)^2 t}\N_{[2,0]}(q,t)+\frac{ q (t+1)^2 (t-1)^2 }{ t (q t-1)^2}\N_{[0,0]}(q,t)\\
       &\quad\quad\quad\quad\quad\quad+\frac{(t-1)(1+q/t)}{1-q}\N_{[1,0]}(q,t)\bigg]\\
       &+p^2\bigg[\scalebox{1.1}{$\frac{ q^2 (t-1)^2 (q t-1)^2}{(q-1)^4 (q+1)^2 t^2}$}\N_{[4,0]}-\scalebox{1.1}{$\frac{q (t-1)^3 (q+t)}{(q-1)^3 t^2}$}\N_{[3,0]}-\scalebox{1.1}{$\frac{(t-1)O}{(q-1)^3 (q+1) t^2 \left(q^3 t-1\right)^2}$}\N_{[2,0]}\\
       &-\scalebox{1.1}{$\frac{(t-1) (q+t) P}{(q-1) t^2 \left(q^2 t-1\right)^2}$}\N_{[1,0]}-\scalebox{1.1}{$\frac{(t-1) Q}{(q-1)^2 (q+1) t^2 (q t-1)^2 \left(q^2 t-1\right)^2}$}\N_{[0,0]}\bigg]+\O(p^3)\Bigg]\,.
    \end{split}
\end{align}
where
\begin{align*}
    O&=q^{10} t^2-3 q^9 t^4+2 q^9 t^3-q^9 t^2-3 q^8 t^5+6 q^8 t^4-q^8 t^2-3 q^7 t^5+6 q^7 t^4-3 q^7 t^3-4 q^7 t^2\\
    &+q^6 t^5-4 q^6 t^4+17 q^6 t^3-13 q^6 t^2+3 q^6 t+q^6+2 q^5 t^5-2 q^5 t^4-2 q^5 t^3-2 q^5 t^2-2 q^5 t\\
    &+2 q^5+q^4 t^5+3 q^4 t^4-13 q^4 t^3+17 q^4 t^2-4 q^4 t+q^4-4 q^3 t^3-3 q^3 t^2+6 q^3 t-3 q^3-q^2 t^3\\
    &+6 q^2 t-3 q^2-q t^3+2 q t^2-3 q t+t^3\,,\\
    P&=q^4 t^3+4 q^3 t^4-4 q^3 t^3-3 q^3 t^2+2 q^3 t+q^3+4 q^2 t^4+2 q^2 t^3-14 q^2 t^2+2 q^2 t+4 q^2+q t^4\\
    &+2 q t^3-3 q t^2-4 q t+4 q+t\,,\\
    Q&=3 q^9 t^4+q^9 t^2-3 q^8 t^6-2 q^8 t^5-q^8 t^4-6 q^8 t^3-2 q^7 t^7+2 q^7 t^6+14 q^7 t^5-2 q^7 t^4-6 q^7 t^3\\
    &+4 q^7 t^2-2 q^7 t-3 q^6 t^7+q^6 t^6+10 q^6 t^5-8 q^6 t^4-3 q^6 t^3+11 q^6 t^2+q^5 t^7+9 q^5 t^6-8 q^5 t^5\\
    &-27 q^5 t^4+10 q^5 t^3+4 q^5 t^2-7 q^5 t+2 q^5-2 q^4 t^7+7 q^4 t^6-4 q^4 t^5-10 q^4 t^4+27 q^4 t^3+8 q^4 t^2\\
    &-9 q^4 t-q^4-11 q^3 t^5+3 q^3 t^4+8 q^3 t^3-10 q^3 t^2-q^3 t+3 q^3+2 q^2 t^6-4 q^2 t^5+6 q^2 t^4+2 q^2 t^3\\
    &-14 q^2 t^2-2 q^2 t+2 q^2+6 q t^4+q t^3+2 q t^2+3 q t-t^5-3 t^3\,.
\end{align*}

\section{Conclusion}\label{sec6}
In this paper, we have established a novel connection between the superconformal index of $\mathcal{N}=4$ $U(N)$ SYM theory and the elliptic Ruijsenaars Schneider classical integrable system. By expressing the index in terms of the kernel function expanded in elliptic Macdonald polynomials, with structure constants $B_\lambda(p,q,t)$ and the orthogonality relation characterized by the elliptic normalization constants $\N_\lambda(p,q,t)$, we have reformulated the index as a concise sum over generalized integral partitions. This representation not only elucidates the underlying integrable structure but also provides a systematic framework for perturbative computations.

A central challenge that remains is the explicit formulas of the constants $\N_\lambda(p,q,t)$ and $B_\lambda(p,q,t)$. While closed expressions for these quantities are currently lacking, we have demonstrated that perturbative expansions in the elliptic parameter $p$ can be obtained by solving the elliptic Ruijsenaars Schneider model order by order. These expansions, combined with the orthogonal basis provided by elliptic Macdonald polynomials, yield a practical method for computing the superconformal index at finite $N$. The results presented for the $N=2$ case illustrate the feasibility of this approach and serve as a template for higher rank generalizations.

Extending our construction to other gauge groups is an interesting
direction. The BCD-type deformed Schur indices suggest that Macdonald-Koornwinder functions and elliptic van Diejen type operators
may play the corresponding role \cite{ren:2025tvx}. Although kernel identities with different balanced conditions for van Diejen operators are known \cite{atai:2022ndn} respect to different type integrals, the
orthogonality and completeness results required to establish a diagonal
expansion analogous to~\eqref{kernelf} are not presently available in the needed
generality. Moreover, the elliptic van Diejen model has eight boundary
parameters, compared with four for Koornwinder polynomials, so such an
expansion may exist only at special parameter values. Whether these
specializations correspond to the $SO$ and $USp$ superconformal indices
remains unclear and is left for future investigation.

Another interest is the study of superconformal indices in the presence of defects. Our method, based on elliptic Macdonald polynomials and their orthogonality, can potentially be adapted to incorporate defect insertions, thereby probing the rich structure of defect CFTs and their holographic duals.

The quantum/classical duality in integrable systems establishes a connection between the Ruijsenaars Schneider classical integrable system and quantum spin chains, as discussed in \cite{gorsky:2013xba, beketov:2015aha, zabrodin:2025byn}. This naturally suggests that the relation between the superconformal index of $\N=4$ $U(N)$ SYM and quantum integrable systems is also a direction worth exploring.

Finally, the formulation of the index as a sum over generalized partitions offers a natural setting for investigating giant graviton expansions. Since the gauge group rank $N$ appears only as a truncation on the length of partitions, the dependence on $N$ is encoded in a simple and transparent way. This opens the door to a systematic study of finite $N$ corrections and the emergence of the large $N$ limit, providing further insights into the AdS/CFT correspondence and the microscopic counting of black hole microstates.

\section*{Acknowledgements}
We thank Xin Wang for helpful discussions. We are supported by the National Natural Science Foundation of China Grants No.12325502 and No.12247103.

\appendix
\section{Coefficients in the elliptic Macdonald polynomials}  \label{appendixA}
The coefficients $A,B,C,D,E,F,G,H,I,J$ appear in the elliptic Macdonald polynomials are list here:
\begin{align}
    A&=q^{11} t^6+q^{11} t^5+q^{11} t^4-3 q^{10} t^6-7 q^{10} t^5+q^{10} t^4-q^{10} t^3-2 q^{10} t^2+2 q^9 t^7+3 q^9 t^6-2 q^9 t^5\notag\\
    &+3 q^9 t^4+3 q^9 t^3+3 q^9 t^2-q^8 t^8-2 q^8 t^7+4 q^8 t^6+12 q^8 t^5-q^8 t^4-7 q^8 t^3+5 q^8 t^2-q^8 t\notag\\
    &+2 q^7 t^7-10 q^7 t^6+5 q^7 t^4-15 q^7 t^3-2 q^7 t^2+2 q^7 t+5 q^6 t^7-9 q^6 t^6-8 q^6 t^5+14 q^6 t^4\notag\\
    &+2 q^6 t^3-4 q^6 t^2+4 q^5 t^6-2 q^5 t^5-14 q^5 t^4+8 q^5 t^3+9 q^5 t^2-5 q^5 t-2 q^4 t^7+2 q^4 t^6+15 q^4 t^5\notag\\
    &-5 q^4 t^4+10 q^4 t^2-2 q^4 t+q^3 t^7-5 q^3 t^6+7 q^3 t^5+q^3 t^4-12 q^3 t^3-4 q^3 t^2+2 q^3 t+q^3-3 q^2 t^6\notag\\
    &-3 q^2 t^5-3 q^2 t^4+2 q^2 t^3-3 q^2 t^2-2 q^2 t+2 q t^6+q t^5-q t^4+7 q t^3+3 q t^2-t^4-t^3-t^2\\
     B&=q^8 t^3+q^8 t^2+q^7 t^3+3 q^7 t^2+q^7 t-q^6 t^4-3 q^6 t^3-2 q^6 t^2-2 q^5 t^3-6 q^5 t^2-3 q^5 t+3 q^4 t^3\notag\\
        &-3 q^4 t+3 q^3 t^3+6 q^3 t^2+2 q^3 t+2 q^2 t^2+3 q^2 t+q^2-q t^3-3 q t^2-q t-t^2-t\\
        C&=q^7 t^4+q^7 t^3-q^7 t^2+2 q^6 t^3-q^6 t^2-2 q^6 t-q^5 t^4-2 q^5 t^3+2 q^5 t^2+q^4 t^4-3 q^4 t^3+q^4 t^2\notag\\
        &+2 q^4 t-2 q^3 t^3-q^3 t^2+3 q^3 t-q^3-2 q^2 t^2+2 q^2 t+q^2+2 q t^3+q t^2-2 q t+t^2-t-1\\
        D&=q^{10} t^5-q^{10} t^4-q^{10} t^3+q^9 t^5+2 q^9 t^4-2 q^9 t^3+q^9 t^2-q^8 t^6-q^8 t^5+q^8 t^4+q^8 t^2+q^8 t\notag\\
        &-2 q^7 t^5-q^7 t^4-q^7 t^2+2 q^6 t^5-2 q^6 t^4+4 q^6 t^3-2 q^6 t+2 q^5 t^5-3 q^5 t^4+3 q^5 t^2-2 q^5 t\notag\\
        &+2 q^4 t^5-4 q^4 t^3+2 q^4 t^2-2 q^4 t+q^3 t^4+q^3 t^2+2 q^3 t-q^2 t^5-q^2 t^4-q^2 t^2+q^2 t+q^2-q t^4\notag\\
        &+2 q t^3-2 q t^2-q t+t^3+t^2-t\\
        E&=q^7 t^4+q^7 t^3-2 q^6 t^3-q^6 t^2-q^6 t-q^5 t^5-q^5 t^4-3 q^5 t^3+4 q^5 t^2+q^5 t+5 q^4 t^4-2 q^4 t^3\notag\\
        &-q^4 t^2+q^4 t-q^4-q^3 t^5+q^3 t^4-q^3 t^3-2 q^3 t^2+5 q^3 t+q^2 t^4+4 q^2 t^3-3 q^2 t^2-q^2 t-q^2\notag\\
        &-q t^4-q t^3-2 q t^2+t^2+t\\
        F&=t^9 q^{20}+t^8 q^{20}-t^{11} q^{19}-3 t^9 q^{19}-t^8 q^{19}-2 t^7 q^{19}-3 t^6 q^{19}+7 t^{10} q^{18}-10 t^9 q^{18}+2 t^8 q^{18}\notag\\
        &+13 t^7 q^{18}+3 t^6 q^{18}+t^5 q^{18}+2 t^4 q^{18}-t^{11} q^{17}+11 t^{10} q^{17}+t^9 q^{17}-7 t^8 q^{17}-6 t^7 q^{17}-2 t^6 q^{17}\notag\\
        &-9 t^5 q^{17}-3 t^4 q^{17}+t^{12} q^{16}-6 t^{11} q^{16}-8 t^{10} q^{16}+2 t^8 q^{16}-4 t^7 q^{16}+13 t^6 q^{16}+21 t^5 q^{16}\notag\\
        &-4 t^4 q^{16}+3 t^3 q^{16}+t^{12} q^{15}+8 t^{10} q^{15}-11 t^9 q^{15}+37 t^8 q^{15}+10 t^7 q^{15}-68 t^6 q^{15}+14 t^5 q^{15}\notag\\
        &-10 t^4 q^{15}-7 t^3 q^{15}-4 t^{11} q^{14}+11 t^{10} q^{14}-51 t^9 q^{14}-10 t^8 q^{14}+52 t^7 q^{14}-37 t^6 q^{14}+7 t^5 q^{14}\notag\\
        &+42 t^4 q^{14}-4 t^3 q^{14}+2 t^2 q^{14}+t^{12} q^{13}-5 t^{11} q^{13}+24 t^{10} q^{13}+17 t^9 q^{13}-22 t^8 q^{13}+32 t^7 q^{13}\notag\\
        &+53 t^6 q^{13}-84 t^5 q^{13}+11 t^4 q^{13}-4 t^3 q^{13}-5 t^2 q^{13}-5 t^{11} q^{12}+5 t^{10} q^{12}+2 t^9 q^{12}-6 t^8 q^{12}\notag\\
        &-98 t^7 q^{12}+61 t^6 q^{12}+17 t^5 q^{12}-26 t^4 q^{12}+28 t^3 q^{12}+t^2 q^{12}+t q^{12}-t^{11} q^{11}+4 t^{10} q^{11}-33 t^9 q^{11}\notag\\
        &+95 t^8 q^{11}-39 t^7 q^{11}-50 t^6 q^{11}+146 t^5 q^{11}-78 t^4 q^{11}-13 t^3 q^{11}+4 t^2 q^{11}-t q^{11}-t^{11} q^{10}\notag\\
        &+14 t^{10} q^{10}-36 t^9 q^{10}+20 t^8 q^{10}+73 t^7 q^{10}-192 t^6 q^{10}+73 t^5 q^{10}+20 t^4 q^{10}-36 t^3 q^{10}+14 t^2 q^{10}\notag\\
        &-t q^{10}-t^{11} q^9+4 t^{10} q^9-13 t^9 q^9-78 t^8 q^9+146 t^7 q^9-50 t^6 q^9-39 t^5 q^9+95 t^4 q^9-33 t^3 q^9\notag\\
        &+4 t^2 q^9-t q^9+t^{11} q^8+t^{10} q^8+28 t^9 q^8-26 t^8 q^8+17 t^7 q^8+61 t^6 q^8-98 t^5 q^8-6 t^4 q^8+2 t^3 q^8\notag\\
        &+5 t^2 q^8-5 t q^8-5 t^{10} q^7-4 t^9 q^7+11 t^8 q^7-84 t^7 q^7+53 t^6 q^7+32 t^5 q^7-22 t^4 q^7+17 t^3 q^7\notag\\
        &+24 t^2 q^7-5 t q^7+q^7+2 t^{10} q^6-4 t^9 q^6+42 t^8 q^6+7 t^7 q^6-37 t^6 q^6+52 t^5 q^6-10 t^4 q^6-51 t^3 q^6\notag\\
        &+11 t^2 q^6-4 t q^6-7 t^9 q^5-10 t^8 q^5+14 t^7 q^5-68 t^6 q^5+10 t^5 q^5+37 t^4 q^5-11 t^3 q^5+8 t^2 q^5+q^5\notag\\
        &+3 t^9 q^4-4 t^8 q^4+21 t^7 q^4+13 t^6 q^4-4 t^5 q^4+2 t^4 q^4-8 t^2 q^4-6 t q^4+q^4-3 t^8 q^3-9 t^7 q^3\notag\\
        &-2 t^6 q^3-6 t^5 q^3-7 t^4 q^3+t^3 q^3+11 t^2 q^3-t q^3+2 t^8 q^2+t^7 q^2+3 t^6 q^2+13 t^5 q^2+2 t^4 q^2\notag\\
        &-10 t^3 q^2+7 t^2 q^2-3 t^6 q-2 t^5 q-t^4 q-3 t^3 q-t q+t^4+t^3\\
        G&=q^{11} t^5-2 q^{10} t^4-q^{10} t^3+q^{10} t^2-q^9 t^6+q^9 t^5+3 q^9 t^4-3 q^9 t^3-q^8 t^5+q^8 t^2+q^8 t-q^7 t^5\notag\\
        &+5 q^7 t^3-3 q^7 t^2-q^7 t+3 q^6 t^5-4 q^6 t^4+q^6 t^3+3 q^6 t^2-q^6 t+q^5 t^5-3 q^5 t^4-q^5 t^3+4 q^5 t^2\notag\\
        &-3 q^5 t+q^4 t^5+3 q^4 t^4-5 q^4 t^3+q^4 t-q^3 t^5-q^3 t^4+q^3 t+3 q^2 t^3-3 q^2 t^2-q^2 t+q^2-q t^4\notag\\
        &+q t^3+2 q t^2-t\\
        H&=-t^{14} q^{34}-t^{13} q^{34}+t^{17} q^{33}-2 t^{16} q^{33}+5 t^{15} q^{33}+6 t^{14} q^{33}-4 t^{13} q^{33}+4 t^{12} q^{33}+6 t^{11} q^{33}\notag\\
        &+t^{17} q^{32}-16 t^{16} q^{32}+3 t^{15} q^{32}+17 t^{14} q^{32}-20 t^{13} q^{32}-25 t^{12} q^{32}-t^{11} q^{32}-3 t^{10} q^{32}-10 t^9 q^{32}\notag\\
        &-t^{18} q^{31}+10 t^{17} q^{31}-19 t^{15} q^{31}-24 t^{14} q^{31}+59 t^{13} q^{31}+32 t^{12} q^{31}-21 t^{11} q^{31}+51 t^{10} q^{31}\notag\\
        &+14 t^9 q^{31}+5 t^7 q^{31}-4 t^{18} q^{30}+6 t^{17} q^{30}+28 t^{16} q^{30}+74 t^{15} q^{30}-95 t^{14} q^{30}-48 t^{13} q^{30}\notag\\
        &+129 t^{12} q^{30}-157 t^{11} q^{30}-73 t^{10} q^{30}+22 t^9 q^{30}-34 t^8 q^{30}-10 t^7 q^{30}-t^{18} q^{29}-19 t^{17} q^{29}\notag\\
        &-115 t^{16} q^{29}+102 t^{15} q^{29}+31 t^{14} q^{29}-320 t^{13} q^{29}+287 t^{12} q^{29}+231 t^{11} q^{29}-134 t^{10} q^{29}\notag\\
        &+94 t^9 q^{29}+102 t^8 q^{29}-12 t^7 q^{29}+10 t^6 q^{29}+t^{19} q^{28}+5 t^{18} q^{28}+68 t^{17} q^{28}-39 t^{16} q^{28}+37 t^{15} q^{28}\notag\\
        &+350 t^{14} q^{28}-77 t^{13} q^{28}-508 t^{12} q^{28}+254 t^{11} q^{28}-74 t^{10} q^{28}-341 t^9 q^{28}+9 t^8 q^{28}-27 t^7 q^{28}\notag\\
        &-36 t^6 q^{28}+t^{19} q^{27}-21 t^{18} q^{27}+8 t^{17} q^{27}-66 t^{16} q^{27}-311 t^{15} q^{27}-359 t^{14} q^{27}+862 t^{13} q^{27}\notag\\
        &-482 t^{12} q^{27}-240 t^{11} q^{27}+810 t^{10} q^{27}+92 t^9 q^{27}-132 t^8 q^{27}+256 t^7 q^{27}+4 t^6 q^{27}+10 t^5 q^{27}\notag\\
        &+t^{19} q^{26}+t^{18} q^{26}+31 t^{17} q^{26}+139 t^{16} q^{26}+636 t^{15} q^{26}-726 t^{14} q^{26}+470 t^{13} q^{26}+933 t^{12} q^{26}\notag\\
        &-1453 t^{11} q^{26}-287 t^{10} q^{26}+682 t^9 q^{26}-655 t^8 q^{26}-186 t^7 q^{26}+33 t^6 q^{26}-47 t^5 q^{26}-9 t^{18} q^{25}\notag\\
        &-22 t^{17} q^{25}-385 t^{16} q^{25}+292 t^{15} q^{25}-425 t^{14} q^{25}-1560 t^{13} q^{25}+1801 t^{12} q^{25}+291 t^{11} q^{25}\notag\\
        &-1042 t^{10} q^{25}+698 t^9 q^{25}+882 t^8 q^{25}-419 t^7 q^{25}+290 t^6 q^{25}+36 t^5 q^{25}+8 t^4 q^{25}+t^{19} q^{24}\notag\\
        &-9 t^{18} q^{24}+120 t^{17} q^{24}-56 t^{16} q^{24}+217 t^{15} q^{24}+1583 t^{14} q^{24}-980 t^{13} q^{24}-525 t^{12} q^{24}\notag\\
        &+1114 t^{11} q^{24}+127 t^{10} q^{24}-2307 t^9 q^{24}+1163 t^8 q^{24}-399 t^7 q^{24}-403 t^6 q^{24}+32 t^5 q^{24}\notag\\
        &-34 t^4 q^{24}-9 t^{18} q^{23}+11 t^{17} q^{23}-88 t^{16} q^{23}-749 t^{15} q^{23}-259 t^{14} q^{23}+540 t^{13} q^{23}-1323 t^{12} q^{23}\notag\\
        &-873 t^{11} q^{23}+3456 t^{10} q^{23}-1331 t^9 q^{23}-815 t^8 q^{23}+1755 t^7 q^{23}-392 t^6 q^{23}+192 t^5 q^{23}\notag\\
        &+28 t^4 q^{23}+5 t^3 q^{23}+17 t^{17} q^{22}+118 t^{16} q^{22}+520 t^{15} q^{22}-518 t^{14} q^{22}+1810 t^{13} q^{22}+1273 t^{12} q^{22}\notag\\
        &-3420 t^{11} q^{22}+396 t^{10} q^{22}+3271 t^9 q^{22}-4224 t^8 q^{22}+1343 t^7 q^{22}-164 t^6 q^{22}-333 t^5 q^{22}\notag\\
        &-16 t^4 q^{22}-15 t^3 q^{22}-2 t^{18} q^{21}+25 t^{17} q^{21}-226 t^{16} q^{21}+353 t^{15} q^{21}-1338 t^{14} q^{21}-1204 t^{13} q^{21}\notag\\
        &+1693 t^{12} q^{21}+638 t^{11} q^{21}-4654 t^{10} q^{21}+5717 t^9 q^{21}-1527 t^8 q^{21}-1028 t^7 q^{21}+1311 t^6 q^{21}\notag\\
        &-158 t^5 q^{21}+117 t^4 q^{21}+9 t^3 q^{21}+2 t^2 q^{21}-3 t^{18} q^{20}+25 t^{17} q^{20}-159 t^{16} q^{20}+560 t^{15} q^{20}\notag\\
        &+494 t^{14} q^{20}-72 t^{13} q^{20}-396 t^{12} q^{20}+4153 t^{11} q^{20}-4714 t^{10} q^{20}-783 t^9 q^{20}+4093 t^8 q^{20}\notag\\
        &-3084 t^7 q^{20}+739 t^6 q^{20}-226 t^5 q^{20}-87 t^4 q^{20}-24 t^3 q^{20}-2 t^2 q^{20}+t^{18} q^{19}+27 t^{17} q^{19}\notag\\
        &-93 t^{16} q^{19}+25 t^{15} q^{19}-302 t^{14} q^{19}+97 t^{13} q^{19}-2900 t^{12} q^{19}+1499 t^{11} q^{19}+4300 t^{10} q^{19}\notag\\
        &-7308 t^9 q^{19}+4980 t^8 q^{19}-859 t^7 q^{19}-658 t^6 q^{19}+374 t^5 q^{19}+86 t^4 q^{19}+43 t^3 q^{19}+2 t^2 q^{19}\notag\\
        &-t^{18} q^{18}-5 t^{17} q^{18}-65 t^{16} q^{18}+69 t^{15} q^{18}-88 t^{14} q^{18}+1311 t^{13} q^{18}+881 t^{12} q^{18}-5207 t^{11} q^{18}\notag\\
        &+8463 t^{10} q^{18}-5799 t^9 q^{18}-861 t^8 q^{18}+3513 t^7 q^{18}-1458 t^6 q^{18}+158 t^5 q^{18}-67 t^4 q^{18}-24 t^3 q^{18}\notag\\
        &-12 t^2 q^{18}+t^{18} q^{17}+6 t^{17} q^{17}+43 t^{16} q^{17}+87 t^{15} q^{17}-256 t^{14} q^{17}-923 t^{13} q^{17}+3478 t^{12} q^{17}\notag\\
        &-6971 t^{11} q^{17}+4099 t^{10} q^{17}+4099 t^9 q^{17}-6971 t^8 q^{17}+3478 t^7 q^{17}-923 t^6 q^{17}-256 t^5 q^{17}\notag\\
        &+87 t^4 q^{17}+43 t^3 q^{17}+6 t^2 q^{17}+t q^{17}-12 t^{17} q^{16}-24 t^{16} q^{16}-67 t^{15} q^{16}+158 t^{14} q^{16}\notag\\
        &-1458 t^{13} q^{16}+3513 t^{12} q^{16}-861 t^{11} q^{16}-5799 t^{10} q^{16}+8463 t^9 q^{16}-5207 t^8 q^{16}+881 t^7 q^{16}\notag\\
        &+1311 t^6 q^{16}-88 t^5 q^{16}+69 t^4 q^{16}-65 t^3 q^{16}-5 t^2 q^{16}-t q^{16}+2 t^{17} q^{15}+43 t^{16} q^{15}+86 t^{15} q^{15}\notag\\
        &+374 t^{14} q^{15}-658 t^{13} q^{15}-859 t^{12} q^{15}+4980 t^{11} q^{15}-7308 t^{10} q^{15}+4300 t^9 q^{15}+1499 t^8 q^{15}\notag\\
        &-2900 t^7 q^{15}+97 t^6 q^{15}-302 t^5 q^{15}+25 t^4 q^{15}-93 t^3 q^{15}+27 t^2 q^{15}+t q^{15}-2 t^{17} q^{14}-24 t^{16} q^{14}\notag\\
        &-87 t^{15} q^{14}-226 t^{14} q^{14}+739 t^{13} q^{14}-3084 t^{12} q^{14}+4093 t^{11} q^{14}-783 t^{10} q^{14}-4714 t^9 q^{14}\notag\\
        &+4153 t^8 q^{14}-396 t^7 q^{14}-72 t^6 q^{14}+494 t^5 q^{14}+560 t^4 q^{14}-159 t^3 q^{14}+25 t^2 q^{14}-3 t q^{14}\notag\\
        &+2 t^{17} q^{13}+9 t^{16} q^{13}+117 t^{15} q^{13}-158 t^{14} q^{13}+1311 t^{13} q^{13}-1028 t^{12} q^{13}-1527 t^{11} q^{13}\notag\\
        &+5717 t^{10} q^{13}-4654 t^9 q^{13}+638 t^8 q^{13}+1693 t^7 q^{13}-1204 t^6 q^{13}-1338 t^5 q^{13}+353 t^4 q^{13}-226 t^3 q^{13}\notag\\
        &+25 t^2 q^{13}-2 t q^{13}-15 t^{16} q^{12}-16 t^{15} q^{12}-333 t^{14} q^{12}-164 t^{13} q^{12}+1343 t^{12} q^{12}-4224 t^{11} q^{12}\notag\\
        &+3271 t^{10} q^{12}+396 t^9 q^{12}-3420 t^8 q^{12}+1273 t^7 q^{12}+1810 t^6 q^{12}-518 t^5 q^{12}+520 t^4 q^{12}+118 t^3 q^{12}\notag\\
        &+17 t^2 q^{12}+5 t^{16} q^{11}+28 t^{15} q^{11}+192 t^{14} q^{11}-392 t^{13} q^{11}+1755 t^{12} q^{11}-815 t^{11} q^{11}-1331 t^{10} q^{11}\notag\\
        &+3456 t^9 q^{11}-873 t^8 q^{11}-1323 t^7 q^{11}+540 t^6 q^{11}-259 t^5 q^{11}-749 t^4 q^{11}-88 t^3 q^{11}+11 t^2 q^{11}\notag\\
        &-9 t q^{11}-34 t^{15} q^{10}+32 t^{14} q^{10}-403 t^{13} q^{10}-399 t^{12} q^{10}+1163 t^{11} q^{10}-2307 t^{10} q^{10}+127 t^9 q^{10}\notag\\
        &+1114 t^8 q^{10}-525 t^7 q^{10}-980 t^6 q^{10}+1583 t^5 q^{10}+217 t^4 q^{10}-56 t^3 q^{10}+120 t^2 q^{10}-9 t q^{10}+q^{10}\notag\\
        &+8 t^{15} q^9+36 t^{14} q^9+290 t^{13} q^9-419 t^{12} q^9+882 t^{11} q^9+698 t^{10} q^9-1042 t^9 q^9+291 t^8 q^9\notag\\
        &+1801 t^7 q^9-1560 t^6 q^9-425 t^5 q^9+292 t^4 q^9-385 t^3 q^9-22 t^2 q^9-9 t q^9-47 t^{14} q^8+33 t^{13} q^8\notag\\
        &-186 t^{12} q^8-655 t^{11} q^8+682 t^{10} q^8-287 t^9 q^8-1453 t^8 q^8+933 t^7 q^8+470 t^6 q^8-726 t^5 q^8\notag\\
        &+636 t^4 q^8+139 t^3 q^8+31 t^2 q^8+t q^8+q^8+10 t^{14} q^7+4 t^{13} q^7+256 t^{12} q^7-132 t^{11} q^7+92 t^{10} q^7\notag\\
        &+810 t^9 q^7-240 t^8 q^7-482 t^7 q^7+862 t^6 q^7-359 t^5 q^7-311 t^4 q^7-66 t^3 q^7+8 t^2 q^7-21 t q^7+q^7\notag\\
        &-36 t^{13} q^6-27 t^{12} q^6+9 t^{11} q^6-341 t^{10} q^6-74 t^9 q^6+254 t^8 q^6-508 t^7 q^6-77 t^6 q^6+350 t^5 q^6\notag\\
        &+37 t^4 q^6-39 t^3 q^6+68 t^2 q^6+5 t q^6+q^6+10 t^{13} q^5-12 t^{12} q^5+102 t^{11} q^5+94 t^{10} q^5-134 t^9 q^5\notag\\
        &+231 t^8 q^5+287 t^7 q^5-320 t^6 q^5+31 t^5 q^5+102 t^4 q^5-115 t^3 q^5-19 t^2 q^5-t q^5-10 t^{12} q^4-34 t^{11} q^4\notag\\
        &+22 t^{10} q^4-73 t^9 q^4-157 t^8 q^4+129 t^7 q^4-48 t^6 q^4-95 t^5 q^4+74 t^4 q^4+28 t^3 q^4+6 t^2 q^4-4 t q^4\notag\\
        &+5 t^{12} q^3+14 t^{10} q^3+51 t^9 q^3-21 t^8 q^3+32 t^7 q^3+59 t^6 q^3-24 t^5 q^3-19 t^4 q^3+10 t^2 q^3-t q^3\notag\\
        &-10 t^{10} q^2-3 t^9 q^2-t^8 q^2-25 t^7 q^2-20 t^6 q^2+17 t^5 q^2+3 t^4 q^2-16 t^3 q^2+t^2 q^2+6 t^8 q+4 t^7 q\notag\\
        &-4 t^6 q+6 t^5 q+5 t^4 q-2 t^3 q+t^2 q-t^6-t^5\\
        I&=t^{10} q^{27}+t^9 q^{27}+3 t^{10} q^{26}-5 t^8 q^{26}-t^7 q^{26}+3 t^{10} q^{25}+t^9 q^{25}-5 t^8 q^{25}+3 t^7 q^{25}+t^6 q^{25}\notag\\
        &+t^5 q^{25}-t^{11} q^{24}-3 t^{10} q^{24}-5 t^9 q^{24}-11 t^8 q^{24}+2 t^7 q^{24}-2 t^{11} q^{23}-2 t^{10} q^{23}-9 t^9 q^{23}+t^8 q^{23}\notag\\
        &+17 t^7 q^{23}+11 t^6 q^{23}-5 t^5 q^{23}+2 t^4 q^{23}-2 t^{11} q^{22}+3 t^{10} q^{22}-17 t^9 q^{22}+13 t^8 q^{22}+8 t^7 q^{22}\notag\\
        &+13 t^6 q^{22}-17 t^5 q^{22}-2 t^4 q^{22}-t^{11} q^{21}+10 t^{10} q^{21}-6 t^9 q^{21}+29 t^8 q^{21}+5 t^7 q^{21}+7 t^6 q^{21}-t^5 q^{21}\notag\\
        &-8 t^4 q^{21}+3 t^3 q^{21}-t^{11} q^{20}+12 t^{10} q^{20}+3 t^9 q^{20}+18 t^8 q^{20}+18 t^7 q^{20}-67 t^6 q^{20}+20 t^5 q^{20}-16 t^4 q^{20}\notag\\
        &+11 t^{10} q^{19}+3 t^9 q^{19}-14 t^8 q^{19}+48 t^7 q^{19}-127 t^6 q^{19}+34 t^5 q^{19}+20 t^4 q^{19}-4 t^3 q^{19}+2 t^2 q^{19}\notag\\
        &+5 t^{10} q^{18}-3 t^9 q^{18}-41 t^8 q^{18}+57 t^7 q^{18}-76 t^6 q^{18}-31 t^5 q^{18}+69 t^4 q^{18}-16 t^3 q^{18}+2 t^2 q^{18}\notag\\
        &+3 t^{10} q^{17}-15 t^9 q^{17}-40 t^8 q^{17}-18 t^7 q^{17}+57 t^6 q^{17}-89 t^5 q^{17}+95 t^4 q^{17}-8 t^3 q^{17}+t^{10} q^{16}\notag\\
        &-34 t^9 q^{16}+23 t^8 q^{16}-81 t^7 q^{16}+164 t^6 q^{16}-28 t^5 q^{16}+16 t^4 q^{16}+4 t^3 q^{16}-10 t^2 q^{16}+2 t^{10} q^{15}\notag\\
        &-38 t^9 q^{15}+75 t^8 q^{15}-89 t^7 q^{15}+52 t^6 q^{15}+110 t^5 q^{15}-95 t^4 q^{15}+12 t^3 q^{15}-10 t^2 q^{15}+t q^{15}\notag\\
        &+t^{10} q^{14}-28 t^9 q^{14}+69 t^8 q^{14}+35 t^7 q^{14}-110 t^6 q^{14}+225 t^5 q^{14}-119 t^4 q^{14}-26 t^3 q^{14}+2 t^2 q^{14}\notag\\
        &-2 t^9 q^{13}+26 t^8 q^{13}+119 t^7 q^{13}-225 t^6 q^{13}+110 t^5 q^{13}-35 t^4 q^{13}-69 t^3 q^{13}+28 t^2 q^{13}-t q^{13}\notag\\
        &-t^{10} q^{12}+10 t^9 q^{12}-12 t^8 q^{12}+95 t^7 q^{12}-110 t^6 q^{12}-52 t^5 q^{12}+89 t^4 q^{12}-75 t^3 q^{12}+38 t^2 q^{12}\notag\\
        &-2 t q^{12}+10 t^9 q^{11}-4 t^8 q^{11}-16 t^7 q^{11}+28 t^6 q^{11}-164 t^5 q^{11}+81 t^4 q^{11}-23 t^3 q^{11}+34 t^2 q^{11}\notag\\
        &-t q^{11}+8 t^8 q^{10}-95 t^7 q^{10}+89 t^6 q^{10}-57 t^5 q^{10}+18 t^4 q^{10}+40 t^3 q^{10}+15 t^2 q^{10}-3 t q^{10}-2 t^9 q^9\notag\\
        &+16 t^8 q^9-69 t^7 q^9+31 t^6 q^9+76 t^5 q^9-57 t^4 q^9+41 t^3 q^9+3 t^2 q^9-5 t q^9-2 t^9 q^8+4 t^8 q^8\notag\\
        &-20 t^7 q^8-34 t^6 q^8+127 t^5 q^8-48 t^4 q^8+14 t^3 q^8-3 t^2 q^8-11 t q^8+16 t^7 q^7-20 t^6 q^7+67 t^5 q^7\notag\\
        &-18 t^4 q^7-18 t^3 q^7-3 t^2 q^7-12 t q^7+q^7-3 t^8 q^6+8 t^7 q^6+t^6 q^6-7 t^5 q^6-5 t^4 q^6-29 t^3 q^6\notag\\
        &+6 t^2 q^6-10 t q^6+q^6+2 t^7 q^5+17 t^6 q^5-13 t^5 q^5-8 t^4 q^5-13 t^3 q^5+17 t^2 q^5-3 t q^5+2 q^5\notag\\
        &-2 t^7 q^4+5 t^6 q^4-11 t^5 q^4-17 t^4 q^4-t^3 q^4+9 t^2 q^4+2 t q^4+2 q^4-2 t^4 q^3+11 t^3 q^3+5 t^2 q^3\notag\\
        &+3 t q^3+q^3-t^6 q^2-t^5 q^2-3 t^4 q^2+5 t^3 q^2-t^2 q^2-3 t q^2+t^4 q+5 t^3 q-3 t q-t^2-t\\
        J&=q^{12} t^4-q^{11} t^2-q^{10} t^5+q^{10} t^4+2 q^{10} t^3-q^{10} t^2-3 q^9 t^4-q^9 t+q^8 t^4-q^8 t^3+2 q^8 t^2\notag\\
        &+q^8 t-3 q^7 t^3+3 q^6 t^4+3 q^6 t-3 q^5 t^2+q^4 t^4+2 q^4 t^3-q^4 t^2+q^4 t-q^3 t^4-3 q^3 t\notag\\
        &-q^2 t^3+2 q^2 t^2+q^2 t-q^2-q t^3+t
\end{align}




\bibliographystyle{JHEP} 
\bibliography{References} 
\end{document}